\documentclass[11pt]{article}
\usepackage[margin=1in]{geometry} 
\UseRawInputEncoding
\usepackage{hyperref}
\usepackage[affil-sl]{authblk}
\usepackage[utf8]{inputenc}
\usepackage[T1]{fontenc}
\usepackage{anyfontsize}
\usepackage{lipsum} 
\usepackage{microtype} 
\usepackage{titlesec} 
\usepackage{graphicx} 
\usepackage{svg}
\usepackage{xcolor}
\usepackage[normalem]{ulem}
\usepackage{marginnote} 

\usepackage{bbm}
\usepackage{xspace}
\usepackage{tabularx}
\usepackage{stmaryrd}
\SetSymbolFont{stmry}{bold}{U}{stmry}{m}{n}
\usepackage{bbm}
\usepackage{url}            
\usepackage{booktabs}       
\usepackage{amsfonts}       
\usepackage{nicefrac}       
\usepackage{microtype}      
\usepackage{amsmath,amsthm}
\usepackage{tikz}
\usetikzlibrary{arrows.meta, positioning}

\usepackage[most]{tcolorbox}
\usepackage{floatrow} 
\usepackage{nicematrix}
\usepackage{natbib}          

\usepackage{newtxmath}
\usepackage{letltxmacro}
\usepackage{bm}
\usepackage{algorithm}
\usepackage[noend]{algpseudocode} 
\usepackage{bbm}
\usepackage{svg}
\usepackage{xcolor}
\usepackage{xspace}
\usepackage{stmaryrd}
\usepackage{bbm}
\usepackage{url}            
\usepackage{booktabs}       
\usepackage{nicefrac}       
\usepackage{microtype}      
\usepackage{letltxmacro}
\usepackage{bm}
\usepackage{caption}
\usepackage{subcaption}
\usepackage{soul}
\usepackage{doi}
\usepackage{fancyvrb,cprotect}
\usepackage{centernot}
\usepackage{fancyhdr}
\graphicspath{ {assets/} }

\usepackage{hyperref} 

\hypersetup{
    colorlinks=true,
    linkcolor=tsutsuji,
    citecolor=kikyou,
    urlcolor=kon-peki
}
\usepackage{multirow}

\usepackage[capitalize]{cleveref}       
\Crefname{claim}{Claim}{Claims}

\usepackage{framed}
\usepackage{rotating}

\usepackage[colorinlistoftodos]{todonotes}

\UseRawInputEncoding
\usepackage{thmtools}
\usepackage[framemethod=TikZ]{mdframed}

\usepackage{marginnote} 
\usepackage{xcolor}     
\usepackage{lipsum}     

\usepackage{enumitem} 

\definecolor{ama-iro}{RGB}{0, 158, 243.0} 
\definecolor{fuyu-gaki}{HTML}{C75146} 
\definecolor{momiji}{RGB}{245, 70, 111} 
\definecolor{hotaru-bi}{RGB}{229,221,58} 
\definecolor{kon-peki}{RGB}{1,120,217} 
\definecolor{shin-kai}{RGB}{77,98,152} 
\definecolor{shin-ryoku}{RGB}{1,145,97} 
\definecolor{yama-budo}{RGB}{171,14,122} 
\definecolor{tsutsuji}{HTML}{C71585} 
\definecolor{mizu}{rgb}{0.0, 0.6, 0.8} 
\definecolor{kikyou}{rgb}{0.4, 0.4, 0.8} 
\definecolor{keshizumi}{RGB}{110, 110, 110} 
\definecolor{usuzumi}{RGB}{140, 140, 140} 

\definecolor{theoremcolor}{RGB}{229,221,58}
\definecolor{definitioncolor}{RGB}{255, 213, 179}
\definecolor{examplecolor}{RGB}{255, 213, 179}

\usepackage{caption}
\newtheoremstyle{theoremstyle}
  {\topsep} 
  {\topsep} 
  {} 
  {} 
  {\bfseries\color{black}} 
  {} 
  {.5em} 
  {\thmname{#1}~\thmnumber{#2}:~\textcolor{black}{\thmnote{[#3]}}} 

\mdfdefinestyle{theoremframe}{
  linecolor=theoremcolor!80,
  linewidth=2pt,
  backgroundcolor=theoremcolor!15,
  roundcorner=5pt,
  nobreak=true,
  topline=false,
  bottomline=false,
  rightline=false,
}
\theoremstyle{theoremstyle}
\newtheorem{theorem}{Theorem}[section]
\surroundwithmdframed[style=theoremframe]{theorem}

\newtheoremstyle{remarkstyle}
  {\topsep} 
  {\topsep} 
  {} 
  {} 
  {\bfseries\color{black}} 
  {} 
  {0.5em} 
  {} 

\mdfdefinestyle{remarkframe}{
  linecolor=momiji!80,
  linewidth=2pt,
  backgroundcolor=momiji!10,
  roundcorner=5pt,
  nobreak=true,
  topline=false,
  bottomline=false,
  rightline=false,
}
\theoremstyle{remarkstyle}
\newtheorem{remark}{Remark}
\surroundwithmdframed[style=remarkframe]{remark}

\newtheoremstyle{definitionstyle}
  {\topsep} 
  {\topsep} 
  {} 
  {} 
  {\bfseries\color{black}} 
  {} 
  {.5em} 
  {\thmname{#1}~\thmnumber{#2}:~\textcolor{black}{\thmnote{[#3]}}} 

\mdfdefinestyle{definitionframe}{
  linecolor=definitioncolor!80,
  linewidth=2pt,
  backgroundcolor=definitioncolor!45,
  roundcorner=5pt,
  nobreak=true,
  topline=false,
  bottomline=false,
  rightline=false,
}
\theoremstyle{definitionstyle}
\newtheorem{definition}[theorem]{Definition}
\surroundwithmdframed[style=definitionframe]{definition}

\newtheoremstyle{lemmastyle}
  {\topsep} 
  {\topsep} 
  {} 
  {} 
  {\bfseries\color{black}} 
  {} 
  {.5em} 
  {\thmname{#1}~\thmnumber{#2}:~\textcolor{black}{\thmnote{[#3]}}} 

\mdfdefinestyle{lemmaframe}{
  linecolor=theoremcolor!80,
  linewidth=2pt,
  backgroundcolor=theoremcolor!05,
  roundcorner=5pt,
  nobreak=true,
  topline=false,
  bottomline=false,
  rightline=false,
}
\theoremstyle{lemmastyle}
\newtheorem{lemma}[theorem]{Lemma}
\newtheorem{corollary}[theorem]{Corollary}
\surroundwithmdframed[style=lemmaframe]{lemma}
\surroundwithmdframed[style=lemmaframe]{corollary}
\newtheorem{claim}[theorem]{Claim}
\surroundwithmdframed[style=definitionframe]{claim}

\newtcolorbox[auto counter, number within=section]{problem}[2][]{colframe=fuyu-gaki, colback=mizu!10, coltitle=white, title={Problem }, sharp corners, boxrule=0.8mm, width=0.99\textwidth, boxsep=2mm, left=3mm, right=3mm, top=2mm, bottom=2mm, breakable}

\newtcolorbox[auto counter, number within=section]{question}[2][]{colframe=fuyu-gaki, colback=mizu!10, coltitle=white, title={Open Question \thetcbcounter: #2}, sharp corners, boxrule=0.8mm, width=0.99\textwidth, boxsep=2mm, left=3mm, right=3mm, top=2mm, bottom=2mm, breakable}

\newcommand{\highlight}[1]{\textcolor{shin-ryoku}{#1}}
\newcommand{\ari}[1]{\textcolor{momiji}{\, \textbf{Ari(Comment):}}\, \textcolor{shin-ryoku}{#1}}
\newcommand{\markComment}[1]{\textcolor{orange}{\textbf{Mark (Comment):}}\textcolor{blue}{\quad#1}}

\newcommand{\clement}[1]{\textcolor{purple}{\textbf{Cl\'ement(Comment):}}\textcolor{ama-iro}{\quad#1}}

\newcommand{\Def}{\overset{\mathtt{def}}{=}}

\makeatletter
\newcommand{\smalldollar}{\mathrel{\mathpalette\small@dollar\relax}}
\newcommand{\small@dollar}[2]{%
  \vcenter{\hbox{%
    $#1\textnormal{\fontsize{0.7\dimexpr\f@size pt}{0}\selectfont\$}$%
  }}%
}
\makeatother

\newcommand{\Size}[1]{\left| #1 \right|}
\newcommand{\RCloseLOpenInterval}[2]{}

\newcommand{\Interval}[2]{\left[\right]}
\newcommand{\OpenInterval}[2]{\left(\right)}

\newcommand{\leftarrowS}{\leftarrow\joinrel\smalldollar}
\newcommand{\xleftarrowS}[1]{\overset{#1}{\leftarrowS}}

\renewcommand{\tilde}[1]{\widetilde{#1}}

\newcommand{\poly}{\operatorname{poly}}
\newcommand{\Naturals}{\mathbb{N}}

\newcommand{\BigO}[1]{O\left(#1\right)}
\newcommand{\BigOTilde}[1]{\tilde{O}\left(#1\right)}

\newcommand{\BigOmega}[1]{\Omega\left(#1\right)}
\newcommand{\BigOmegaTilde}[1]{\tilde{\Omega}\left(#1\right)}
\newcommand{\BigTheta}[1]{\Theta\left(#1\right)}

\newcommand{\SmallOmega}[1]{\omega\left(#1\right)}

\newcommand{\Yes}{\mathsf{Yes}}
\newcommand{\No}{\mathsf{No}}

\newcommand{\TV}[2]{d_{\text{TV}}\left(#1,#2\right)}
\newcommand{\RL}[2]{\texttt{RL}\left(#1,#2\right)}
\newcommand{\EMD}[2]{\mathrm{EMD}\left(#1,#2\right)}
\newcommand{\Property}{\Pi}
\newcommand{\Domain}{[N]}
\newcommand{\DomainSize}{N}
\newcommand{\samples}{\leftarrowS} 
\newcommand{\iidSamples}{\xleftarrowS{\text{i.i.d}}}

\newcommand{\Dist}{\mathcal{D}}
\newcommand{\DistPrime}{\textcolor{black}{\mathcal{D}'}}

\newcommand{\Uniform}[1]{\mathsf{Uniform}\left[#1\right]}
\newcommand{\DistSet}[1]{\Delta\left(#1\right)}
\newcommand{\Support}[1]{\text{Supp}\left({#1}\right)}
\newcommand{\SupportSize}[1]{\left|\Support{#1}\right|}

\newcommand{\Prob}[1]{\Pr\left[#1 \right]}
\newcommand{\PProb}[2]{\Pr_{#2}\left[#1 \right]}

\newcommand{\Claimed}[1]{\textcolor{usuzumi}{\widetilde{#1}}}
\newcommand{\ClaimedDist}[1]{\Claimed{\Dist}\left[#1\right]}
\newcommand{\Tag}[1]{\Claimed{\BucketOf{\Proximity}{#1}}}
\newcommand{\True}[1]{\textcolor{black}{#1}}
\newcommand{\TrueDist}[1]{{\True{\Dist}\left[#1\right]}}

\newcommand{\NumBuckets}{L_\Proximity}

\newcommand{\BucketOf}[2]{\text{Bucket}_{#1}(#2)}
\newcommand{\HistFactor}{\tau'}
\newcommand{\TrueHist}{\{\True{p_j}\}_{j \in \BucketIndices}}

\newcommand{\ClaimHistExpanded}{\{\Claimed{p_j}\}_{j \in \BucketIndices}}
\newcommand{\BucketIndices}{[\NumBuckets]}

\newcommand{\ApproxHistParam}{(\DomainSize, \Proximity)}

\newcommand{\Reject}{\textcolor{black}{\mathsf{Reject}}}
\newcommand{\Fail}{\textcolor{black}{\mathsf{FAIL}}}
\newcommand{\Accept}{\textcolor{black}{\mathsf{Accept}}}
\newcommand{\Protocol}{\Pi}
\newcommand{\TesterFunc}{\mathsf{T}}
\newcommand{\tester}[1]{\TesterFunc^{\left(#1\right)}}
\newcommand{\Tester}[2]{\tester{#1}\left(#2\right)}
\newcommand{\prover}{\mathsf{P}}
\newcommand{\Prover}[1]{\prover^{#1}}
\newcommand{\chProver}{{\widetilde{\prover}}}
\newcommand{\ChProver}[1]{\chProver^{#1}}
\newcommand{\ProofSystem}[2]{\Pi\left(\Prover{#1},\tester{#1}; #2\right)}
\newcommand{\ChProofSystem}[2]{\Pi\left(\ChProver{#1},\tester{#1}; #2\right)}

\newcommand{\outputs}[1]{\text{out}\left[#1\right]}

\newcommand{\Proximity}{\tau}

\newcommand{\Cond}{\texttt{Cond}}
\newcommand{\PCond}{\texttt{PCond}}
\newcommand{\ICond}{\texttt{ICond}}
\newcommand{\Samp}{\texttt{Samp}}
\newcommand{\BPPTester}{\mathsf{BPP}}

\newcommand{\IPTester}{\mathsf{IP}}

\newcommand{\yStar}{{y^\star}}
\newcommand{\yStarJ}{{y_{j}^\star}}
\newcommand{\jStar}{{j^\star}}

\newcommand{\NeighbourhoodExpanded}[2]{U^{\Dist}_{#2}\left(#1\right)}

\newcommand{\NeighbourhoodAlpha}[1]{\NeighbourhoodExpanded{#1}{\alpha}}

\title{Interactive Proofs For Distribution Testing With Conditional Oracles}

\author{%
  \begin{minipage}{0.24\linewidth}\centering
    Ari Biswas\\
    {\small University of Warwick}
  \end{minipage}\hfill
  \begin{minipage}{0.24\linewidth}\centering
    Mark Bun\\
    {\small Boston University}
  \end{minipage}\hfill
  \begin{minipage}{0.24\linewidth}\centering
    Cl\'ement Canonne\\
    {\small University of Sydney}
  \end{minipage}\hfill
  \begin{minipage}{0.24\linewidth}\centering
    Satchit Sivakumar\\
    {\small Boston University}
  \end{minipage}
}

\date{}

\begin{document}

\maketitle
\begin{abstract}
We revisit the framework of interactive proofs for distribution testing, first introduced by Chiesa and Gur (ITCS 2018), which has recently experienced a surge in interest, accompanied by notable progress (e.g., Herman and Rothblum, STOC 2022, FOCS 2023; Herman, RANDOM~2024). 
In this model, a data-poor verifier  determines whether a probability distribution has a property of interest by interacting with an all-powerful, data-rich but untrusted prover bent on convincing them that it has the property. While prior work gave sample-, time-, and communication-efficient protocols for testing and estimating a range of distribution properties, they all suffer from an inherent issue: for most interesting properties of distributions over a domain of size $N$, the verifier must draw at least $\Omega(\sqrt{N})$ samples of its own. While sublinear in $N$, this is still prohibitive for large domains encountered in practice.

In this work, we circumvent this limitation by augmenting the verifier with the ability to perform an exponentially smaller number of more powerful (but reasonable) \emph{pairwise conditional} queries, effectively enabling them to perform ``local comparison checks'' of the prover's claims.
We systematically investigate the landscape of interactive proofs in this new setting, giving polylogarithmic query and sample protocols for (tolerantly) testing all \emph{label-invariant} properties, thus demonstrating exponential savings without compromising on communication, for this large and fundamental class of testing tasks.
\end{abstract}

\tableofcontents

\section{Introduction}\label{sec:intro}

 Distribution testing, as introduced by~\citet{BatuFRSW00}, is a mature subfield of property testing~\citep{GoldreichGR98, rubinfeldRobust} aimed at investigating statistical properties of an unknown distribution given sample access to it. 
 Given a property (a set of distributions) and a proximity parameter $\Proximity \in (0, 0.1]$, distribution testing algorithms output $\Accept$ if the distribution is in the property (or close to it), or $\Reject$ if the distribution is $\Proximity$-far from the property, both with high probability. Closeness and farness are quantified with respect to a prespecified notion of distance, typically total variation distance.
The primary motivation behind distribution testing is to design testing algorithms for deciding properties with sample complexity sub-linear in the domain size $\DomainSize$ (which is demonstrably more efficient than learning the distribution, which requires  drawing $\BigTheta{\DomainSize}$ samples).
Accordingly, over the last two decades, researchers have extensively studied the sample complexity of numerous distribution properties, such as simple uniformity testing~\citep{goldreich2011testing} (testing whether a distribution is uniform over its entire domain), support size decision problem~\citep{RaskhodnikovaRSS09,ValiantV11,YihongP19,PintoH25} (testing whether a distribution's support is within some pre-specified range), 
and many more: see, e.g.,~\cite[Chapter~11]{Goldreich17} and~\cite{Rubinfeld12,Canonne20,CanonneTopicsDT2022} for a more thorough introduction to distribution testing. 
Unfortunately, although distribution testing is often more efficient than learning the distribution, it is still prohibitively expensive for practical use. 
For example, it is known that generalized uniformity testing (testing whether a distribution is uniform over its support) over a domain of size $\DomainSize$ requires $\BigOmega{\DomainSize^{2/3}}$ samples~\citep{BatuC17,DiakonikolasKS18}, which can be impractical for large domain sizes. 
Even simple uniformity testing requires $\Omega(\sqrt{N})$ samples~\citep{Paninski08}, and its \emph{tolerant} testing version (which asks to distinguish distributions \emph{close} to uniform from those which are far) needs $\BigOmega{\DomainSize/\log \DomainSize}$ samples~\citep{valiant2017estimating}.

In the face of these limitations, a nascent line of work \citep{chiesa2018proofs, herman2022verifying, herman2023doubley, herman2024public} has asked a related question: \textit{with testing being hard by itself, what is the complexity of \emph{verifying} the properties of a distribution given sample access to it?}
Here, in addition to drawing samples from the distribution, the tester is allowed to interactively communicate with an omniscient but \emph{untrusted} prover that knows the distribution in its entirety. 
The idea here is to leverage the provers extra knowledge about the distribution, with the hope that checking the provers' claims is easier than naively testing the property.
While this model of verifiable computation has only recently been explored in the context of distribution testing, it has been an active area of research in other areas of theoretical computer science for over 40 years (see for e.g. \citep{goldwasserZK1985, silvioSoundProofs, rvw14, goldwasserMuggles15, berman2018, arun2024jolt}).
It models settings where a centralized organization (for example, a company turning billions of dollars of profit) has the ability to collect large amounts of data and learn distributions to high precision, while end-users may not have the same ability. 
At the same time, the company might have incentives to lie, and so verifying whether the company is being truthful is important in this setting.
The work of \citet{chiesa2018proofs} shows that the verification of \emph{any} distribution property over domain $\Domain$ can be reduced to identity testing\footnote{Identity testing refers to the task of testing if a distribution is exactly equal to a pre-specified reference distribution or is $\Proximity$-far from it.}, with communication \emph{superlinear} in the domain size. 
Follow up work \citep{herman2022verifying, herman2023doubley} recovers this result for the broad class of \emph{label-invariant properties}, while only requiring communication \emph{sub-linear} in the domain size.
More specifically, the work of \citet{herman2022verifying, herman2023doubley} show that for label-invariant properties, verification requires only $\BigO{\sqrt{\DomainSize}}$ samples and $\BigOTilde{\sqrt{\DomainSize}}$ communication, even though, as mentioned earlier, testing some properties in this class could require $\Theta(N/\log N)$ samples. 
Here, a property is \emph{label-invariant} (also known as \emph{symmetric}) if the names of the elements themselves are not significant to the decision outcome (see Definition \ref{defn:label-invariant-prop}). 
Testing if a distribution is uniform over its support (also known as generalized uniformity testing) is an example of a label-invariant property. 

Unfortunately, while a significant improvement over unaided testing, requiring $\BigO{\sqrt{\DomainSize}}$ samples from the verifier can still be prohibitive when considering massive domains. 
Further, there is a matching sample complexity lower bound -- verification of even basic label-invariant properties such as checking if a distribution is uniform over its entire domain requires $\Omega(\sqrt{N})$ samples. 
To summarise: For most properties, with access to \emph{only} samples from a distribution, it is impossible for any tester to do better than drawing $\BigOmega{\sqrt{\DomainSize}}$ samples, with or without the help of a prover.
To bypass these limitations and develop more practical algorithms, in this work we study  verifiers that can make a very small number of calls to a more powerful \emph{conditional sampling} oracle. These oracles were introduced in the context of distribution testing \citep{chakraborty2013power, CRS:14}; allowing the tester to condition that samples from the oracle come from a subset $S$ of the domain, of their choosing. 
The oracle responds with a sample with probability re-normalised over $S$.
If no element in $S$ is supported, the oracle responds with $\Fail$.
Since specifying an arbitrary set may considered be unrealistic for practical purposes, a commonly studied restriction is the pairwise conditional sampling model ($\PCond$), where the specified sets are restricted to be of size exactly $2$ or the entire domain (thus, just a regular sample from the distribution). 
These oracles can be thought of as allowing for local comparisons between the probabilities of two points. 
While access to a $\PCond$ oracle can be significantly helpful for problems like simple uniformity testing, it is unclear from prior work whether it results in more efficient testing for the general class of label-invariant properties. 
Verification with access to a $\PCond$ oracle (or any type of conditional sampling oracle) has also, to the best of our knowledge, not been explored. 
In our quest to find practical algorithms that work for large domains, we thus ask the following question.

\begin{framed}
        {\it Can label-invariant properties be verified in a (query, sample and communication)-efficient  way when the tester has access to a $\PCond$ oracle? 
        }
\end{framed}

\subsection{Our Results}

Our main result is an \emph{exponential} query complexity separation between testing and verification for testing label-invariant properties with access to a $\PCond$ oracle. A detailed accounting of our results and comparison to existing work can be found in Table~\ref{tab:comparison}. A description follows.

 \begin{table}[h!]
\centering
\begin{tabularx}{\linewidth}{l *{4}{>{\centering\arraybackslash}X} >{\centering\arraybackslash}X >{\centering\arraybackslash}X}
\toprule
 & Query Complexity Without Prover & Query Complexity With Prover & Communication & Rounds \\
\midrule
  $\Samp$  & $\BigOmegaTilde{\frac{\DomainSize}{\log \DomainSize}}$ \cite{valiant2017estimating} 
         & $\BigOTilde{\sqrt{\DomainSize}}$ \citep{herman2022verifying,herman2023doubley}
         & $\BigOTilde{\sqrt{\DomainSize}}$ \citep{herman2022verifying,herman2023doubley}
         & 2 \citep{herman2022verifying,herman2023doubley} \\\\
$\PCond$ & $\bm{\BigOmega{\DomainSize^{1/3}}}$ (Theorem~\ref{cor:lowerbound})
         & $\bm{\text{poly}(\log \DomainSize,\frac{1}{\Proximity})}$
         & $\bm{\BigOTilde{\sqrt{\DomainSize}}}$ 
         & $\bm{\text{poly}(\log \DomainSize, \frac{1}{\Proximity})}$   \\
         &&\multicolumn{3}{c}{(Theorem~\ref{thm:main-thm})}\\
\bottomrule
\end{tabularx}
\caption{Results on testing and verifying label-invariant properties under different types of access to the distribution. We state the best known lower bounds for label invariant properties.
For $\Samp$ the lower bound is for entropy estimation, whereas for $\PCond$ it is the support size decision problem described in Section \ref{sec:suppsize}.
The upper bounds apply for all label-invariant properties.
Our results are highlighted in bold.
}  
\label{tab:comparison}
\end{table}

One might have initially hoped that the power of a $\PCond$ oracle allows us to test label-invariant properties efficiently, even without the help of a prover. Indeed, with access to the full power of the $\Cond$ model (where arbitrary subsets $S$ can be queried and a sample conditional on $S$ is obtained), \cite{chakraborty2013power} show that this class of properties over a domain of size $N$ can be tested with $O(\poly \log N)$ queries to the $\Cond$ oracle. 
Our first result dashes this hope~--~we show a lower bound on the number of $\PCond$ queries required to test label-invariant properties with constant proximity parameter $\Proximity$, demonstrating that the $\PCond$ oracle is not much better in the worst case than the sampling oracle for this class of properties. 
Specifically, we show that a simple variant of the support size distinguishing problem for distributions over a domain of size $\DomainSize$ requires $\Omega(\DomainSize^{1/3})$ queries to a $\PCond$ oracle (the exact same as with access to only a sampling oracle).
Thus, unaided, there exist (label-invariant) properties for which the $\PCond$ oracle is not much better than just sample access. 

\begin{theorem}[Informal Version of \Cref{cor:lowerbound}]\label{thm:informallb}

  There exists a label-invariant property $\Property$ such that every tester with access to a $\PCond$ oracle for $\Property$ with proximity parameter  $\Proximity \leq 1/2$ and failure probability  $0.01$ must make $\BigOmega{\DomainSize^{1/3}}$ queries.
\end{theorem}

The above lower bound motivates the investigation of verification with access to a $\PCond$ oracle. 
As mentioned earlier, \cite[Proposition 3.4]{chiesa2018proofs} showed that with super-linear communication complexity, there exists a reduction from verification to identity testing.
Instantiating this reduction with an identity tester using $\PCond$ oracles \citep[Theorem 1.5]{narayanan2020distribution}, we get that there exists an interactive proof system for every property with super-linear communication complexity that makes only $O(\sqrt{\log N}/\Proximity^2)$ queries to the $\PCond$ oracle. 
However, super-linear communication is also prohibitive for practical algorithms; the proof systems by \citet{herman2022verifying, herman2023doubley} require the prover to only communicate $\tilde{O}(\sqrt{\DomainSize})$ domain elements, but still achieve the sample complexity of identity testing (for the class of label-invariant properties).
Could we also hope to achieve such communication while maintaining similar query complexity as that of identity testing? 
The main result of this paper is an affirmative answer to this question.
Specifically, we give an interactive proof system for tolerantly verifying \emph{any} label-invariant property that has communication complexity $\tilde{O}(\sqrt{N})$ and query complexity $\poly(\log N)$ (suppressing the dependence on the proximity parameter).

\begin{theorem}[Informal Label-Invariant Tolerant Verification Theorem (Theorem~\ref{thm:main-thm})]\label{thm:introlabelinv}
  Fix a label-invariant property $\Property$ over a domain $[N]$ and proximity parameters $\Proximity_c, \Proximity_f \in (0, 1/2]$. 
  There exists a polylogarithmic (in $\DomainSize$) round interactive protocol $\Protocol$ between an honest verifier $\TesterFunc$, and an omniscient untrusted prover $\Prover{\Dist}$, where the verifier has $\PCond$ access to $\Dist$, such that at the end of the interaction the verifier satisfies the following conditions:
	\begin{enumerate}
    \item{\textbf{Completeness:} If the prover follows the protocol as prescribed, and $\TV{D}{\Property} \leq \Proximity_c$, then
		      \[ \Prob{\outputs{\ProofSystem{\Dist}{\Proximity_c, \DomainSize}} = \Accept } \geq 2/3 \]
		      }
        \item{\textbf{Soundness:} If $\TV{D}{\Property} \geq \Proximity_f$, then for any prover $\tilde{\Prover{\Dist}}$ 

		      \[ \Prob{\outputs{\ChProofSystem{\Dist}{\Proximity_c, \DomainSize}} = \Reject } \geq 2/3 \]

		      }
	\end{enumerate}

The complexity of the verifier is as follows:
  \begin{enumerate}
    \item \textbf{Query Complexity + Sample Complexity}: $O\left(\poly(\log N, 1/(\Proximity_f - \Proximity_c) )\right)$
    \item \textbf{Communication Complexity:} $\tilde{O} \left( \sqrt{N} \poly(1/(\Proximity_f - \Proximity_c)) \right)$
  \end{enumerate}
\end{theorem}

In the process of proving this result, we give protocols for more basic primitives that may be of independent interest. Most significantly, we give an interactive proof system that is able to calculate the approximate probability mass of any point\footnote{Provided it does not have prohibitively small probability mass in its neighborhood.} in the domain using communication complexity $\tilde{O}(\sqrt{N})$ and query complexity $O(\poly(\log N, 1/\Proximity))$. 
As we will explain in the techniques section to follow, this is a key technical workhorse in our protocol for verifying label-invariant properties.

\begin{theorem}[Informal Version of \Cref{lemma:approximate-single}]\label{informal:approxsingle}
    Fix a domain $\Domain$. For every $\Proximity > 0$ and $\delta \in (0,1/2)$, there exists a $O(\ \poly(\log N, \log 1/ \delta, \frac{1}{\tau}))$-round  interactive proof system such that the verifier $\TesterFunc$ with access to a $\PCond$ oracle satisfies the following.
	\begin{enumerate}

		\item \textbf{Completeness:} For every distribution $\Dist$, if the prover $\Prover{\Dist}$ is honest, then
		      \[ \Prob{ \TesterFunc \text{ outputs } (\yStar, \ClaimedDist{\yStar}) \text{ s.t } \frac{\ClaimedDist{\yStar}}{\TrueDist{\yStar}} \in \left[\frac{1}{(1+\Proximity)} , 1 + \Proximity \right] } \ge 1 - \delta \]

		\item{ \textbf{Soundness:} For any cheating prover $\ChProver{\Dist}$, then

		      \[ \Prob{ \TesterFunc \text{ outputs } \Reject \lor \TesterFunc \text{ outputs } (\yStar, \ClaimedDist{\yStar}) \text{ s.t } \frac{\ClaimedDist{\yStar}}{\TrueDist{\yStar}} \in [1/(1 + \Proximity)^2, (1 + \Proximity)^2]} \ge 1-\delta \]

		      }

	\end{enumerate}
The complexity of the verifier is as follows:
  \begin{enumerate}
    \item \textbf{Query Complexity + Sample Complexity}: $O\left(\poly(\log N, 1/(\Proximity) )\right)$
    \item \textbf{Communication Complexity:} $\tilde{O} \left( \sqrt{N} \poly(1/(\Proximity)) \right)$
  \end{enumerate}

\end{theorem}

\subsection{Technical Overview}

In this section, we give an overview of our protocol for verifying label-invariant properties.

\paragraph{Unlabelled Bucket Histogram:} There is now a long line of work on the testing and verification of label-invariant properties \citep{BatuFRSW00, valiantthesis, chakraborty2013power, herman2022verifying, herman2023doubley}, and a key object used in this work is the unlabelled \emph{approximate} $\Proximity$-bucket histogram of a distribution. 
Bucketing corresponds to partitioning the interval $[0,1]$ into smaller multiplicative probability intervals (see Definition \ref{defn:bucketting-or-partition}). 
The $\Proximity$-bucket histogram divides the interval $[0,1]$ into $\BigOTilde{\log \DomainSize /\Proximity}$ buckets where the $\ell$\textsuperscript{th} bucket is a set of domain elements with individual probability mass in the range $(\Proximity(1+\Proximity)^{\ell} / N, \Proximity(1+\Proximity)^{\ell+1} / N]$. 
The \emph{approximate} unlabelled bucket histogram of a distribution then corresponds to a list of $\tilde{O}(\log N / \Proximity)$ fractions, where the $\ell$\textsuperscript{th} element of the list is the fraction of domain points whose probability lies in the range specified by the $\ell$\textsuperscript{th} bucket (see Definition \ref{defn:approx-hist}).
It is well known (see \citep{valiantthesis}) that the \emph{approximate} unlabelled $\Proximity$-bucket histogram of a distribution is a sufficient statistic to (tolerantly) test any label-invariant property with proximity parameter(s) $\BigO{\Proximity}$. 
Thus, similar to prior work \citep{herman2022verifying, herman2023doubley, herman2024public}, our protocol also focuses on efficiently verifying an unlabelled $\Proximity$-bucket histogram given to us by the prover.

\paragraph{Using Pairwise Comparisons to Learn Bucket Histogram:} Note that the unlabelled bucket histogram is a distribution over the buckets of the $\Proximity$-bucket histogram and is hence a distribution over a domain of size $\tilde{O}(\log \frac{N}{\Proximity})$. 
Hence, by standard results in distribution learning, $\tilde{O}(\log N/\Proximity^2)$ samples from this bucket distribution would be sufficient to learn it. 
However, sampling from this bucket distribution is non-trivial since a sample from the original distribution $\Dist$ does not come with information about histogram bucket index. 

While sampling from the bucket distribution might be hard with $\Samp$ access to the distribution, one might be more optimistic about the possibility of sampling from the $\Proximity$-approximate histogram with $\PCond$ queries. 
In particular, one approach that we might take is the following: the verifier draws a dataset of size $\tilde{O}(\log N/\Proximity^2)$ (enough to learn the histogram), and sends these samples to the prover, who responds with the bucket index of each sample.
From how a $\Proximity$-histogram is defined, if $x$ and $y$ belong to buckets $i$ and $j$ respectively, then this implies that the ratio of the probability masses of $x$ and $y$ under $\Dist$ is guaranteed to be in the interval $\left[\frac{1}{(1+\Proximity)^{|j-i|}}, (1 + \Proximity)^{|j-i|}\right]$.
As $\PCond$ access allows us to conditionally sample from a set restricted to two domain elements, it allows us to approximately learn the ratio of their probability masses up to a multiplicative constant (see Lemma \ref{lemma:pcond-empirical-guarantee}).
Equipped with this power, for each pair $x \neq y$ from the set of drawn samples, the verifier uses the $\PCond$ oracle to check if the learned ratios align with the provers claims.
If the prover were to lie significantly, then for at least one pair of samples, the claimed ratio would significantly different from the learned ratio.
Unfortunately, this simplistic strategy comes with two pitfalls. 
Firstly, assuming the above strategy was sound, naively comparing elements with arbitrary bucket indices could require $\BigOmega{\DomainSize}$ $\PCond$ queries if the elements being compared had significantly different probability masses
(see the \nameref{lemma:pcond-empirical-guarantee}, wherein $K$ could be as large as $\DomainSize$).
This would be as bad as learning the distribution itself. 
Secondly, and more importantly, the above strategy is \emph{not} sound. 
It does not catch a prover that ``slides'' all samples into different buckets \emph{in the same way}, i.e., it lies about \emph{every} bucket index by the same offset.
As an example, consider two distributions: $\Dist_1$ is the uniform distribution over $N^{1/4}$ domain elements and $\Dist_2$ is the uniform distribution over $\sqrt{N}$ domain elements. 
The bucket histograms of both involve a unit mass on a single (but different) bucket. 
However, pairwise comparisons between samples taken from either distribution will always reveal a ratio that is approximately $1$, since the probabilities of all elements in the support of both distributions are identical. 
Hence, given distribution $\Dist_1$, the prover can output the bucket histogram of distribution $\Dist_2$, and it is impossible for a verifier to catch it purely by using the test described above. 

A remedy to these obstacles lies in the following  observation.
If we had a good estimate of the probability mass of \emph{a single point} $y$ in the domain, then we could resolve the soundness issue discussed above.
We simply use the $\PCond$ oracle to learn the ratio of probabilities between each of our samples and $y$.
Using this ratio, and knowledge of the approximate mass of $y$, we can compute estimates of the probabilities of all samples.
This would squash the sliding attack described above. 
To deal with the first issue (that of the probability mass of $y$ being very far from that of a sample), we would need to do more than learn just one value of $y$.
Instead, we could learn the mass of a point $y_j$, for every bucket $j$ that has large enough mass.
This way, for any sample $x$ in the tester's set of samples (which are likely to come from buckets with sufficiently large mass), we can find some $y_j$ that is in a near-by bucket with high probability.

\paragraph{Verifying the probability of points:} Given that we simply need to identify the probability of a few points in the domain, one might expect that this could be done even without access to a prover~---~this would give a query-efficient tester for label-invariant properties. 
However, this ends up being a surprisingly challenging task. Indeed, our lower bound in Theorem~\ref{thm:informallb} shows that this is impossible (if we wanted to bypass the $\poly(\DomainSize)$ lower bound). 
This indicates a power of an untrusted prover; it is able to certify the probability of a few points in the distribution support. 
Indeed, the proof system with super-linear communication \cite{chiesa2018proofs} achieves something stronger- it certifies the entire distribution. 
Since we only need to certify the mass of at most $\BigO{\log \DomainSize}$ points, we ask if this be done in a more communication efficient way?

\paragraph{Support Size Verification:} Inspired by the ``sliding'' cheating prover from earlier, we consider the orthogonal but related problem of verifying the support size of a flat distribution.\footnote{A distribution is \emph{flat} if it is uniform over its support.} 
We will subsequently show that a protocol for this problem can be combined with the ability of a $\PCond$ oracle to learn neighbourhoods around points, enabling us to solve the probability approximation problem for ``relevant'' domain elements.

Given a support size claim represented by four numbers $A',A,B,B'$, corresponding to the claim that $A' < A \leq \Support{\Dist} \leq B < B'$, our hope is to accept if the claimed support size range is accurate, and reject if the true support is larger than $B'$ or smaller than $A'$. 
Given different values of $A$ and $B$, we develop a number of tests to verify with \emph{sub-linear} communication, the support size assuming the distribution is uniform over its support\footnote{We relax this condition in the main protocol, but assuming uniformity makes the description more intuitive.}. 
We summarize the ideas below.

If the claimed support size upper bound $B$ is small (that is, $\BigO{\sqrt{\DomainSize}}$), we could ask the prover to send us the claimed support of the distribution. 
If the prover lies, and the true support is actually much larger, then taking a few samples from the distribution would give us a domain element outside the claimed support, thus catching the lie of the prover. 
If the true support is much smaller than $A$, then taking a number of uniform samples from the claimed support sent by the prover would result in a sample outside the support of the distribution, which could be easily detected with a few $\PCond$ queries (see Lemma \ref{lemma:isinsupport}). 
This gives us a protocol with a constant number of queries, and communication complexity roughly $O(B)$. 

On the other hand, if the claimed support size lower bound $A$ is large (that is, $\SmallOmega{\sqrt{\DomainSize}}$), then asking the prover to send the support is not communication-efficient. Our approach instead involves using uniform samples from the domain. 
The first test is to catch provers that lie that the support is much larger than it really is. 
It involves drawing $O(N/A)$ uniform samples $S_1$ from the domain, and sending them to the prover. 
We ask the prover to send back a sample in this subset that is in the support of the distribution. 
If the true support is much smaller than $A$, there are (with high probability) no samples in the support of the distribution in $S_1$, and we can ensure the prover does not cheat by checking whether any element it sends back is in the support using a constant number of $\PCond$ queries. 
The second test catches provers that lie that the support is much smaller than it really is. 
It involves drawing $O(N/B')$ samples $z_1,\dots,z_m$ uniformly from the domain and permuting them with one sample $x$ taken from the distribution. 
We ask the prover to identify the index of $x$ in the permuted set.
If the true support is smaller than $B$, then w.h.p, we expect that none of $z_1,\dots,z_s$ are in the support of the distribution and hence, the honest prover can identify $x$ exactly. 
On the other hand, if the support is larger than $B'$, we expect that at least one of $z_1,\dots,z_s$ is in the support of the distribution, and the prover is unable to tell what the sample inserted by the prover was (since it could be $x$ or $z_i$).  
This gives us a protocol with a constant number of queries, and communication complexity roughly $O(N/A)$. 
Balancing parameters in these tests to optimize the communication complexity, we get an overall protocol for support size verification with communication complexity roughly $\sqrt{N}$.

We emphasize that the above outline is a simplification of the truth, and sweeps important details under the rug. 
Recall that our main goal is to certify the histogram of a distribution. 
In an attempt to do so, we will use the support size protocol above repeatedly as a sub-routine. 
This requires bounding the soundness and completeness errors in a meaningful way.
As described above, these protocols do not have low enough soundness and completeness error to be used as sub-routines. 
Additionally, the above description assumes that distributions are exactly flat. 
In practice this will not be the case, and we need to be able to handle distributions where the probability ratio between any two elements in the support is upper bounded by some constant $\alpha$ (nearly flat).
In Section \ref{sec:suppsize}, we show how to analyse the protocols above to facilitate amplification of soundness and completeness errors, and make the protocol work for the more general class of $\alpha$-flat distributions. 

\paragraph{Estimating Probability of a Point using Support Size Verification:} Finally, we explain how we can use the support size protocols to estimate the probability of a point. 
Prior work \citep{canonne2015testing} using the $\PCond$ oracle has shown that it can be used to estimate the mass within a multiplicative $(1+\Proximity)$-neighbourhood of any point\footnote{As long as the point does not have prohibitively small probability mass under the distribution.} (see \nameref{lemma:estimate-neighborhood}).
We ask the prover to send us a point $\yStar$ \footnote{Recall that in the final protocol, we will ask the prover to send us points from every bucket with sufficiently large mass. For simplicity, we consider a single point in this description.} with sufficiently large mass in its neighbourhood (by an averaging argument, at least one histogram bucket needs to have $\Proximity/\log N$ mass, ensuring that such a point exists, see Claim \ref{claim:exists_y_star}), and tell us the bucket the $\yStar$ belongs to.
We then use the $\PCond$ oracle to estimate the mass within the multiplicative neighbourhood of $\yStar$. 
The learned mass of the neighborhood divided by the prover's claimed probability mass\footnote{The bucket index gives a lower and upper bound on the true probability mass of $\yStar$. This is sufficient to catch a cheating prover.} of $\yStar$ gives us bounds on the number of elements in $\yStar$'s neighbourhood.
Additionally, by definition the neighbourhood of $\yStar$ will be nearly flat.  
Hence, we have reduced the problem to a claim about the support size of a nearly-uniform distribution over a subset of the domain. 
Observe that we can sample from the distribution restricted to this bucket using $\PCond$ queries---since there is sufficient mass in the neighbourhood of the point, $O(\poly \log N)$ samples from the distribution will contain at least one sample from the bucket, and we can use $\PCond$ queries between the samples and $\yStar$ to find out which sample it is (see \nameref{lemma:sampling-lemma}). 
If the prover was telling the truth (or rather did not lie egregiously), then the support size claim holds and the verifier will accept, thereby giving us a point and its approximate probability mass\footnote{The claimed mass of $\yStar$ is derived from the bucket sent by the prover.}. 
If the prover significantly lied, then the support size claim will be false and the prover will be caught. 
We note that the final analysis needs to handle some additional subtleties, since the $\PCond$ oracle is itself only approximate and does not provide perfect comparisons (which can affect the sampling), and additionally the bucket boundaries and the neighborhood of the provided point do not overlap precisely. 
We refer to Sections~\ref{sec:onebucklearner} and~\ref{sec:label-invariance} for the details.
Once we have estimated the probability of important points, we can use the ratio learning techniques discussed earlier to certify the prover's claim about the bucket histogram of the distribution. 

\subsection{Other Related Work:}

\paragraph{Interactive Proofs for Distribution Testing:} As mentioned earlier, interactive proofs for verifying distribution properties were first introduced in the work of \cite{chiesa2018proofs}. Follow-up work \citep{herman2022verifying, herman2023doubley} studied interactive proofs for verifying \emph{label-invariant} properties, focusing on sublinear communication, and doubly efficient protocols (i.e. computationally-efficient and sample-efficient generation of a proof). \cite{herman2024public} studied \emph{public-coin }interactive proofs for testing label-invariant properties, where the verifier has no private randomness. \cite{HermanR24} gives communication-efficient and sample-efficient interactive proofs for more general distribution properties that can be decided by uniform polynomial-size circuits with bounded depth. \cite{HermanR25} introduces \textit{computationally sound} interactive proofs and shows communication-, time-, and sample-efficient protocols for any distribution property that can be decided in polynomial time given explicit access to the full list of distributiom probabilities. All of the above works assume that the verifier only has sample access to the distribution, and the verifier sample complexity in all of them is $\Omega(\sqrt{N})$ (where $N$ is the size of the domain). 

\paragraph{Interactive Proofs for Learning:} A related but orthogonal line of work \citep{GoldwasserRSY21, MutrejaS23, GurJKRSS24, CaroHINS24, CarosHINS25} focuses on interactive proofs for verifying learning problems-  for a specific hypothesis class $H$, given access to an untrusted prover, the goal is for a verifier to output an accurate hypothesis from $H$ for an underlying unknown distribution $D$ if the resource-rich prover is honest, and to reject if the prover lies egregiously. Different types of resource asymmetry between the prover and the verifier are explored in these papers -- including differing number of samples, different computational complexities, different types of access to the underlying function (sample vs query), and differing access to computational resources (classical vs quantum computation and communication).

\paragraph{Distribution testing under conditional oracles:} Our work focuses on interactive proofs for \textit{verifying} distribution properties under conditional sampling models. There is a long line of work on \emph{testing} with access to conditional samples. \citet{chakraborty2013power, CRS:14} introduced the conditional sampling ($\Cond$) model and its more restricted variants ($\PCond, \ICond$). They gave algorithms for uniformity testing, tolerant uniformity testing, identity testing etc. in these conditional sampling models with query and sample complexity significantly better than the $\Samp$ model. Follow-up work shows improved bounds for identity testing (and its tolerant version), tolerant uniformity testing, and new algorithms for other tasks such as equivalence testing and support size problem in the $\Cond$ model \citep{falahatgar2015faster, narayanan2020distribution, ChakrabortyKM23, ChakrabortyCK24}. 
There is also a line of work studying the power of non-adaptive queries in the conditional sampling model \citep{AcharyaCK15, KamathT19}. The $\PCond$ model was studied in detail by \cite{narayanan2020distribution} who gave optimal bounds for identity testing and tolerant uniformity testing in this model, improving on results from \cite{canonne2015testing}. Testing under other types of conditional sampling has also been studied in the literature including subcube conditioning, where the distribution is supported on the hypercube, and the tester is allowed to ask for conditional samples from subcubes \citep{BhattacharyyaC18, CanonneCKLW21, ChenJLW21, KumarMM23, ChakrabartyCR0W25}, and coordinate conditional sampling \citep{BlancaCSV25} (a version of subcube conditioning where all but one coordinate is fixed to a specific configuration and a sample is obtained from the remaining coordinate). Testing under other types of access to the distribution such as Probability Mass Function queries or Cumulative Distribution Function queries has also been studied in the literature \citep{BatuDKR02, RubinfeldS05, GuhaMV09,  CanonneR14,OnakS18}. 

\section{Preliminaries} \label{sec:prelims}

\paragraph{General Notation}
We use $\DistSet{\mathcal{X}}$ to denote the set of all probability distributions over some set $\mathcal{X}$, and $[N]$ as shorthand for the set $\{1, \dots, N\}$.
We use the notation $x \samples \Dist$ to indicate $x$ was sampled according to $\Dist$.
Given a set $S \subseteq \mathcal{X}$, we use $\Dist[S] \Def \sum_{y \in S} \Dist[y] \Def \sum_{y \in S} \PProb{x = y}{x\samples \Dist}$ to denote the probability of hitting set $S$ when sampling from $\Dist$, and $\Dist[y]$ is the probability of seeing $y$ when sampling according to $\Dist$.
When we say distance of a distribution $\Dist$ from a property $\Property \subseteq \DistSet{\mathcal{X}}$, we mean $\TV{\Dist}{\Property} \Def \min_{\DistPrime \in \Property}\TV{\Dist}{\DistPrime}$, where $\TV{\Dist}{\Dist'}$ denotes the total variation distance between two distributions.
We use $\Support{\Dist}$ to denote the support of a distribution  $\Dist$.
For any set $\mathcal{X}$, we use $\mathcal{S}_{\mathcal{X}}$ to denote the set of permutations over the set.
Throughout, we use tilde notation $\Claimed{\cdot}$ with lighter font to denote an untrusted prover's claims.
For example, we use $\ClaimedDist{x}$ to denote a prover's claim about the probability mass of $x \in \Domain$ under $\Dist$, and $\TrueDist{x}$ as the true mass.
Similarly, $\Tag{x}$ will denote the provers claim about the histogram bucket index (see Definition \ref{defn:bucketting-or-partition}), versus $\BucketOf{\Proximity}{x}$ which denotes the true index under $\Dist$.

\begin{definition}[Label-Invariant Properties]\label{defn:label-invariant-prop}
	A property $\Property \subseteq \DistSet{\mathcal{X}}$ is said to be \emph{label-invariant} (or \emph{symmetric}) if it is closed under permutations of the domain: that is, if $\Dist \in \Property$ and $\pi \in \mathcal{S}_{\mathcal{X}}$, then the distribution $\Dist_\pi$ defined by
	\[
    \Dist_\pi[x] \Def \Dist[\pi(x)], \qquad x\in \mathcal{X},
	\]
	is also in $\Pi$.
\end{definition}

The set of distributions over $\Domain$ with support size less than 20 is an example of a label-invariant property.
Next, we give the definition for the \emph{true} histogram of a distribution $\Dist$, followed by the $\Proximity$-approximate histogram, which can be viewed as a discretisation of the true histogram of a distribution. 
For label-invariant properties it makes more sense to use the following notion of distance instead of total variation distance.

\begin{definition}[Relabelling Distance]Given two distributions $\Dist, \Dist' \in \DistSet{\Domain}$, we define the relabelling distance between $\Dist$ and $\Dist'$ as 

  \[ \RL{\Dist}{\Dist'} \Def \min_{\pi \in \mathcal{S}_{\Domain}} \TV{\Dist}{\Dist'_{\pi}} =  \min_{\pi \in \mathcal{S}_{\Domain}} \frac{1}{2}\sum_{x \in \Domain} \left| \Dist[x] - \Dist'[\pi(x)]\right| \]
  
\end{definition}

\begin{definition}[Histograms Of Distributions]
  Define the normalised \emph{histogram} of $\Dist \in \DistSet{\Domain}$ as the non-negative function $h_\Dist: [0,1] \rightarrow [0, 1]$
\[
  h_{\Dist}(z) \;=\; \frac{1}{\DomainSize}\bigl|\{i \in \Domain : \TrueDist{i} = z\}\bigr|
  \quad \text{for each } z \in [0,1].
\]

\end{definition}

Thus, $h_\Dist(z)$ counts the fraction of domain elements with probability mass $z$. 
For a function any non-negative function $h$, we use $\Support{h}$ to denote the set $\{x \in [0, 1]: h(x) > 0 \}$. 
Observe that 
\[
  \int_{0}^1 h_\Dist(z)\, dz = 1.
\]

\begin{definition}[$\ApproxHistParam$-Bucketing]
	\label{defn:bucketting-or-partition}
  Fix $\Dist \in \DistSet{\Domain}$ and $\Proximity \in (0,1]$.  
  Let $\NumBuckets = \left\lceil\frac{\log(\DomainSize/\Proximity)}{\log(1+\Proximity)}\right\rceil = \BigO{\log(\DomainSize/\Proximity)}$ denote the number of buckets.	 
  We denote with sets $\{ B_j^{\Dist}\}_{j\in \BucketIndices}$ the $\Proximity$-partitioning (bucketing) of the elements in the support of $\Dist$ where
	\[
  B_\ell^{(\Dist, \Proximity)} = \left\{x \in \Domain :  \frac{\Proximity(1+\Proximity)^{\ell-1}}{\DomainSize}  < \Dist[x]  \leq \frac{\Proximity(1+\Proximity)^{\ell}}{\DomainSize} \right\} \text{ for all } \ell \in \{1, \ldots, \NumBuckets\}	
  \]
  and 
  	\[
  B_0^{(\Dist, \Proximity)} = \left\{x \in \Domain :  \Dist[x] \le \frac{\Proximity}{\DomainSize} \right\} 
  \]
 
    where $\BucketIndices = \left\{0, 1, \dots, L \right\}$. 
\end{definition}
Throughout this document, for any $x \in \Domain$, we use $\BucketOf{\Proximity}{x} \in \BucketIndices$ to denote the bucket to which $x$ belongs.
When the context is clear, we will often drop the superscript in $B_j^{(\Dist, \Proximity)}$, and write just $B_j$.
Note that the buckets above are disjoint, and cover the entire $[0,1]$ interval. 
So for any $z \in [0,1]$ it can belong to exactly \emph{one} bucket. This allows to give the following definition for an approximate histogram.

\begin{definition}[Approximate Histogram of a Distribution]\label{defn:approx-hist}
  Given a $(\DomainSize, \Proximity)$-bucketing of a distribution $\Dist$:  $\left\{ B_j\right\}_{j \in \BucketIndices}$, we define the normalised $\Proximity$-histogram of $\Dist$ with the following non-negative function $h_\Dist^{(\Proximity)}: [0,1] \to [0, 1]$, where  
\[
  h_\Dist^{(\Proximity)}(z) \;=\;
  \begin{cases}
    \TrueDist{B_0}, & \text{if } z \le \frac{\Proximity}{\DomainSize}, \\
    \TrueDist{B_\ell}, & \text{if } z \in \left(\frac{\Proximity(1+\Proximity)^{\ell-1}}{\DomainSize},\,\frac{\Proximity(1+\Proximity)^{\ell}}{\DomainSize}\right], \; \ell \in \{1, \dots, \NumBuckets\}.
  \end{cases}
\]

\end{definition}

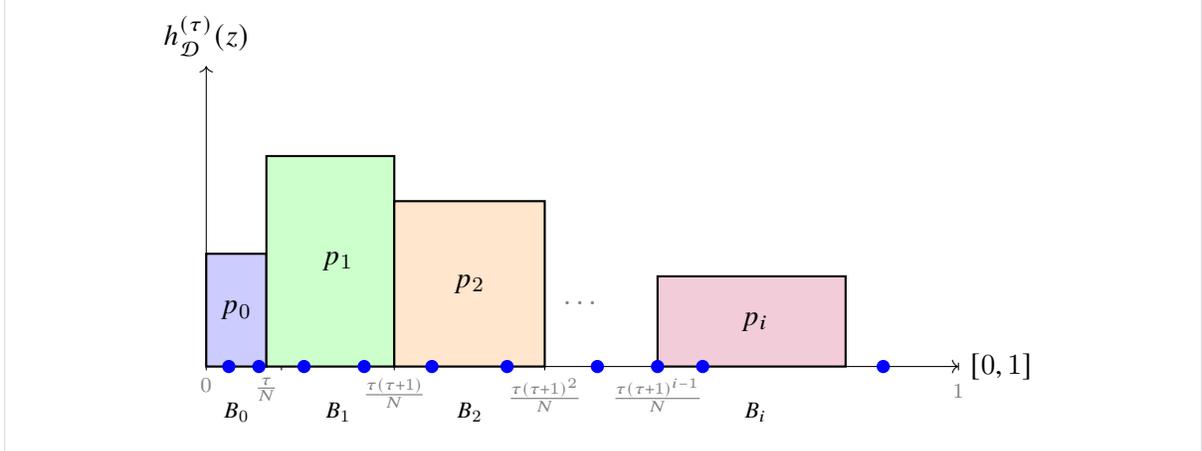
\begin{figure}[h]
\centering
\fcolorbox{gray!30}{white}{%
\begin{minipage}{0.95\textwidth}
\centering
\begin{tikzpicture}[scale=1.0]
\pgfmathsetmacro{\one}{0.8}
\pgfmathsetmacro{\bucketlabels}{0.8}
\pgfmathsetmacro{\intervaly}{-0.02}
\pgfmathsetmacro{\buckety}{-0.2}
\draw[->] (0,0) -- (10,0) node[right] {$[0,1]$};
\draw[->] (0,0) -- (0,4) node[above] {$h_\Dist^{(\Proximity)}(z)$};
\node[below, gray] at (0,\intervaly) {\scriptsize $0$};
\node[below, gray] at (\one,\intervaly) {\scriptsize $\frac{\Proximity}{\DomainSize}$};
\node[below, gray] at (2.5,\intervaly) {\scriptsize $\frac{\Proximity(\Proximity+1)}{\DomainSize}$};
\node[below, gray] at (4.5,\intervaly) {\scriptsize $\frac{\Proximity(\Proximity+1)^2}{\DomainSize}$};
\node[below, gray] at (5, 1) {$\cdots$};
\node[below, gray] at (6,\intervaly) {\scriptsize $\frac{\Proximity(\Proximity+1)^{i-1}}{\DomainSize}$};
\node[below] at (8, 1) {$\cdots$};
\node[below, gray] at (10,-0.1) {\scriptsize $1$};
\draw (0,-0.05) -- (0,0.05);
\draw (1,-0.05) -- (1,0.05);
\draw (2.5,-0.05) -- (2.5,0.05);
\draw (4.5,-0.05) -- (4.5,0.05);
\draw (10,-0.05) -- (10,0.05);
\draw[fill=blue!20, draw=black, thick] (0,0) rectangle (\one,1.5);
\draw[fill=green!20, draw=black, thick] (\one,0) rectangle (2.5,2.8);
\draw[fill=orange!20, draw=black, thick] (2.5,0) rectangle (4.5,2.2);
\draw[fill=purple!20, draw=black, thick] (6,0) rectangle (8.5,1.2);
\node at ({\one/2},0.75) {$\True{p_0}$};
\node at (1.75,1.4) {$\True{p_1}$};
\node at (3.5,1.1) {$\True{p_2}$};
\node at (7.3,0.6) {$\True{p_i}$};
\node[below=5pt, scale=\bucketlabels] at ({\one/2},\buckety) {$B_0$};
\node[below=5pt, scale=\bucketlabels] at (1.75,\buckety) {$B_1$};
\node[below=5pt, scale=\bucketlabels] at (3.5,\buckety) {$B_2$};
\node[below=5pt, scale=\bucketlabels] at (7.3,\buckety) {$B_i$};
\fill[blue] (0.3,0) circle (2.5pt);
\fill[blue] (0.7,0) circle (2.5pt);
\fill[blue] (1.3,0) circle (2.5pt);
\fill[blue] (2.1,0) circle (2.5pt);
\fill[blue] (3.0,0) circle (2.5pt);
\fill[blue] (4.0,0) circle (2.5pt);
\fill[blue] (5.2,0) circle (2.5pt);
\fill[blue] (6.0,0) circle (2.5pt);
\fill[blue] (6.6,0) circle (2.5pt);
\fill[blue] (9.0,0) circle (2.5pt);
\end{tikzpicture}
\vspace{0.5em}
\end{minipage}%
}
\caption{ The approximate histogram $h_\Dist^{(\Proximity)}$ induced by a $(\DomainSize, \Proximity)$-partitioning of $\Domain$ according to $\Dist$.
The blue dots denote domain elements $x \in \Domain$ positioned on the interval $[0,1]$ according their probability mass $\TrueDist{x}$.
Buckets $(B_1, \ldots, B_L)$ denote a disjoint $\Proximity$-partitioning of domain, where $B_i$ is the set of domain elements $x$, such that $\TrueDist{x} \in \left(\frac{\Proximity(1+\Proximity)^{i-1}}{\DomainSize},\,\frac{\Proximity(1+\Proximity)^{i}}{\DomainSize}\right]$.
The mass of bucket $i$ is denoted with $\True{p_i} = \sum_{x \in B_i} \TrueDist{x}$.}
\label{fig:approximate-hist-schematic}
\end{figure}

Figure \ref{fig:approximate-hist-schematic} illustrates how the $(\DomainSize, \Proximity)$ bucketing of the domain a distribution induces an approximate histogram.
Observe in that above definition, the domain of $h_\Dist^{(\Proximity)}$ has at most $\NumBuckets + 1$ values i.e. we have discretised the true histogram into $\NumBuckets +1$ values. 
Thus, the above histogram can be succinctly represented with the following collection of items: $\TrueHist$, where for all $j \in \BucketIndices$
  \[ 
    \True{p_j} \Def \TrueDist{B_j^{(\Dist, \Proximity)}} \Def \sum_{x \in B_j^{(\Dist, \Proximity)}} \TrueDist{x} 
  \]

In the rest of the document, we use this succinct notation to represent approximate histograms of a distribution.
As histograms are defined by non-negative functions over $[0,1]$, it is possible to define distances between histograms with the Earth-Mover Distance Function.

\begin{definition}[Earth-Mover Distance and Relative Earth-Mover Distance]
Let $h$ and $h'$ be two non-negative functions over $[0,1]$ such that
\[
  \sum_{x \in \Support{h}} h(x) \;=\; \sum_{x \in \Support{h'} } h'(x).
\]
The \emph{Earth-Mover Distance (EMD)} between $h$ and $h'$, denoted
$\EMD{h}{h'}$, is defined as
\[
  \EMD{h}{h'} \;=\; \min_{m} \;
\sum_{x \in S(h)} \sum_{y \in S(h')} m(x,y)\cdot |x-y|,
\]
where the minimum is taken over all non-negative functions
$m: \Support{h} \times \Support{h'} \to \mathbb{R}$ such that:
\begin{align*}
\sum_{y \in \Support{h}} m(x,y) &= h(x), &\forall x \in \Support{h}, \\
\sum_{x \in \Support{h'}} m(x,y) &= h'(y), &\forall y \in \Support{h'}.
\end{align*}

\end{definition}

The mental model, and the source of the metrics nomenclature is the following: Imagine we start with a non negative function $h$ which has a hole for each $z \in [0,1]$, filled with $h(z)$ volume of dirt. 
In the case of approximate histograms, the holes with dirt are the buckets with non-zero mass, and the support size of $h$ is the number of buckets.
Given a histogram $h'$, the earth-mover distance between $h$ and $h'$ is the minimum volume of dirt that must be moved from the support of $h$ to the support of $h'$ to transform $h$ into $h'$.
The following lemma is due to \cite[Lemma 5]{goldreich2020relation} links distances between histograms to re-labelling distances. 
\begin{lemma}[Relationship Between Re-labelling Distance And Earth Mover Distance]\label{lemma:rel-rl-emd}For any two distributions $\Dist, \Dist' \in \DistSet{\Domain}$, let $h_\Dist$ and $h_\Dist'$ denote the histogram for $\Dist$ and $\Dist'$ respectively. 
  Then we have 
  \[
    \RL{\Dist}{\Dist'} = \frac{1}{2}\EMD{h_\Dist}{h_\Dist'} 
  \]
  
\end{lemma}

\paragraph{Oracle Access Models}
    Testers, provers, and verifiers in this paper take as input parameters such as the data domain size $N$, error tolerance, and completeness and soundness failure probabilities. They are also given oracle access to the distribution $\Dist$ comprising the problem instance as follows.
	A conditional oracle for distribution $\Dist \in \DistSet{\Domain}$, denoted by $\Cond$, takes as input a set $S \subseteq \Domain$ with the constraint $\Dist(S) > 0$, and outputs an $x \in S$ with probability $\Dist_S[x] = \frac{\TrueDist{x}}{\TrueDist{S}}$.
	If the testing algorithm sends $S$ such that $\Dist(S) = 0$, then the oracle outputs $\Fail$.
	A $\PCond$ oracle for a distribution $\Dist \in \DistSet{\DomainSize}$ is a conditional oracle with the restriction that inputs to the oracle must be of size exactly two or $\DomainSize$.

The round complexity counts the number of rounds of interaction between the prover and the verifier, where one round consists of a message from the verifier, \emph{and} the prover's response.
For any algorithm $\mathsf{A}$, we use parentheses $(\cdot)$ to distinguish between $\mathsf{A}^\Dist$ having full knowledge of the distribution, and $\mathsf{A}^{(\Dist)}$ only being able to access $\Dist$ via oracles. 
Before describing proof systems, we recap the definition of a tolerant testing algorithm \emph{without} a prover.

\begin{definition}[Tolerant $\BPPTester$ Tester]
	A $\BPPTester$ tolerant tester for property $\Property \subseteq
		\DistSet{\Domain}$ is a testing algorithm $\Tester{\Dist}{\DomainSize, \Proximity_c, \Proximity_s} $ such that for any $\Proximity_c, \Proximity_s \in (0, 1]$ we have

	\begin{itemize}
		\item{\textbf{Completeness}: For every $\Dist \in \DistSet{\Domain}$ such that $\TV{\Dist}{\Property} \leq \Proximity_c$,  \[ \Prob{\Tester{\Dist}{\DomainSize, \Proximity} = \Yes} \geq \frac{2}{3}\]}

		\item{\textbf{Soundness}: For every $\Dist \in \DistSet{\Domain}$ such that $\TV{\Dist}{\Property} > \Proximity_s$,  we have
		      \[
			      \Prob{\Tester{\Dist}{\DomainSize, \Proximity} = \Yes} \leq \frac{1}{3}
		      \]}
	\end{itemize}

\end{definition}

%
%
%
%
%
%
%


\paragraph{Proof Systems} The complexity of a proof system is measured by the number of oracle calls a verifier makes. 
In this work, to make accounting more granular, we report independent and identically distributed (iid) samples from the distribution and $\PCond$ queries separately.
The total query complexity is the sum of the two measures.
The communication complexity is measured in the number of bits\footnote{As any domain element can be represented in $\log_2 \DomainSize$ bits, when suppressing log factors in asymptotic expressions, the communication complexity can be read as the number of domain elements exchanged.} that the prover and verifier exchange throughout the proof. 

\begin{definition}[Tolerant $\IPTester$ Tester]
	A tolerant $\IPTester$ for property $\Property \subseteq
		\DistSet{\Domain}$ is testing algorithm $\Tester{\Dist}{\Proximity_c, \Proximity_s, \DomainSize}$ that interactively exchanges messages with an omniscient but untrusted prover $\Prover{\Dist}(\Proximity_c, \Proximity_s, \DomainSize)$, such that at the end of the interaction, for any $\Proximity_c, \Proximity_s \in (0, 1]$ we have

	\begin{itemize}
		\item{\textbf{Completeness}: For every $\Dist \in \DistSet{\Domain}$, such that $\TV{\Dist}{\Property} \leq \Proximity_c$, there exists a prover $\prover$ such that

		      \[
			      \Prob{\ProofSystem{\Dist}{\DomainSize, \Proximity} = \Yes} \geq \frac{2}{3}
		      \]

		      }

		\item{\textbf{Soundness}: For every $\Dist \in \DistSet{\Domain}$ such that $\TV{\Dist}{\Property} > \Proximity_s$, for \emph{every} prover strategy $\chProver$  we have

		      \[
			      \Prob{\ChProofSystem{\Dist}{\DomainSize, \Proximity} = \Yes} \leq \frac{1}{3}
		      \]

		      }
	\end{itemize}

\end{definition}

Throughout this document, a distribution that is uniformly distributed over its support is referred to as a flat distribution.

\begin{definition}[Nearly Flat Distributions]
	\label{def:nearly:flat}
	Given a parameter $\kappa \geq 1$, a probability distribution $\Dist$ over $\Domain$ is said to be \emph{$\kappa$-flat} if the probabilities of any two elements of its support are within a factor $\kappa$; that is,
	\[
		\max_{x\in\Support{\Dist}} \Dist[x] \leq \kappa\cdot\min_{x\in\Support{\Dist}} \Dist[x]\,.
	\]
	(Note that $1$-flat distributions are exactly the flat distributions.) We further refer to $2$-flat distributions as \emph{nearly flat}.
\end{definition}

\subsection{Useful Tools}

In this section, we describe a couple of useful lemmas that we repeatedly use later in the document.

\begin{lemma}[Compare Algorithm]\label{lemma:pcond-empirical-guarantee}
	There is an algorithm \texttt{COMPARE} which, provided $\PCond$ oracle for an arbitrary probability distribution $\Dist \in \DistSet{\Domain}$, has the following guarantees. On input $\gamma, \beta \in (0,1]$ and $K\geq 1$, as well as two distinct elements $x,y \in \Domain$, the algorithm makes $\BigO{\frac{K\log 1/\beta}{\gamma^2}}$ queries to the oracle and outputs either $\textsf{High}$, $\textsf{Low}$, or a value $\alpha > 0$, such that the following holds.
    \begin{itemize}
        \item If $\frac{\Dist[x]}{\Dist[y]} > K$, then with probability at least $1-\beta$ the algorithm outputs either $\textsf{High}$, or a value $\alpha > 0$ such that $\left|\frac{\Dist[x]}{\Dist[y]} - \alpha \right| \leq \gamma$;
        \item If $\frac{\Dist[x]}{\Dist[y]} < 1/K$, then with probability at least $1-\beta$ the algorithm outputs either $\textsf{Low}$, or a value $\alpha > 0$ such that $\left|\frac{\Dist[x]}{\Dist[y]} - \alpha \right| \leq \gamma$;
        \item Otherwise, with probability at least $1-\beta$ it outputs a value $\alpha > 0$ such that $\left|\frac{\Dist[x]}{\Dist[y]} - \alpha \right| \leq \gamma$.
    \end{itemize}
\end{lemma}

The proof follows from the guarantees of \citep[Lemma~2]{canonne2015testing}, specialized for the case of $\PCond$.

\paragraph{Checking Membership in Support}

\begin{algorithm}
	\caption{\texttt{IsInSupport} algorithm.}
	\label{alg:pcond-support-checker}
	\begin{algorithmic}[1]
		\Require Given $x \in \Domain$, and $y \in \Domain$, failure probability parameter $\beta\in (0,1)$,  oracle access to $\PCond^{\Dist}$ for a nearly flat distribution $\Dist$.
		\Ensure Check if  $y \in \Support{\Dist}$
		\State Set $T = \lceil \log_{3/2}(1/\beta)\rceil$
		\ForAll{$1\leq t\leq T$}
		\State $s\samples \PCond^{\Dist}(\{x,y\})$ \label{step:querytopcond}
		\If{$s = y$}
		\Return $\Yes$
		\EndIf
		\EndFor
		\State \Return $\No$
	\end{algorithmic}
\end{algorithm}

\begin{lemma}
	\label{lemma:isinsupport}
  The algorithm $\texttt{IsInSupport}$ has the following guarantees. On inputs $x,y \in \Domain$ and $\beta\in(0,1)$, it makes $\BigO{\log(1/\beta)}$ queries to the $\PCond^{\Dist}$ oracle. Further, if $\Dist$ is nearly flat(Definition \ref{def:nearly:flat}), then
	\begin{itemize}
		\item If $y\in \Support{\Dist}$, it returns $\Yes$ with probability at least $1-\beta$;
		\item If $y\notin \Support{\Dist}$ and $x \in \Support{\Dist}$, it returns $\No$ with probability $1$.
	\end{itemize}
\end{lemma}
\begin{proof}
	The claimed query complexity is immediate from our setting of $T$, observing that the only $\PCond$ queries are on Line~\ref{step:querytopcond}. Turning to the correctness, assume that $\Dist$ is nearly flat.

	If $x=y$, the claims are trivially true and vacuous, respectively; we can thus assume $x\neq y$. Then, on the one hand, if $y\in \Support{\Dist}$, then (regardless of whether $x$ is in $\Support{\Dist}$ or not), $\Dist[y] \geq \frac{1}{2}\cdot \Dist[x]$, and so
	\[
		\Prob{\PCond^{\Dist}(\{x,y\}) = x} = \frac{\Dist[x]}{\Dist[x]+\Dist[y]} \leq \frac{2}{3}\,.
	\]
	The probability that none of the $T$ independent queries returned $y$ is then at most $2/3)^T \leq \beta$.
	On the other hand, if $y\notin \Support{\Dist}$ but $x \in \Support{\Dist}$, then $\Prob{\PCond^{\Dist}(\{x,y\}) = x} = 1$, and so the algorithm returns $\No$ with probability 1.
\end{proof}

\section{Lower Bounds For Pairwise Conditional Testers}
In this section, we give a label-invariant property over a domain $[N]$ where any $\PCond$-tester needs $\poly(N)$ queries. To do this, we will start by defining testing for promise problems and over meta-distributions.

\begin{definition}[Testing for Promise Problem]
	A $\BPPTester$ tester for property promise $\Property^{\Yes}, \Property^{\No} \subseteq
		\DistSet{\Domain}$ is a testing algorithm $\TesterFunc$ such that for any failure probability $\beta \in (0, 1]$ we have

	\begin{itemize}
		\item{\textbf{$\Yes$ Case}: For every $\Dist \in \Property^{\Yes}$,  \[ \Prob{\Tester{\Dist}{\DomainSize, \beta} = \Yes} \geq 1-\beta\]}

		\item{\textbf{$\No$ Case}}: For every $\Dist \in \Property^{\No}$,  we have
		      \[
			      \Prob{\Tester{\Dist}{\DomainSize, \beta} = \No} \geq 1-\beta
		      \] 
	\end{itemize}
\end{definition}

\begin{definition}[Testing for Meta-Distributions]
	Given distribution subsets $\Property^{\Yes}, \Property^{\No} \subseteq
		\DistSet{\Domain}$, and distributions $Q_{\Yes}$ and $Q_{\No}$ over $\Property^{\Yes}, \Property^{\No}$ respectively, a $\BPPTester$ tester for $Q_{\Yes}$, $Q_{\No}$ is a testing algorithm $\TesterFunc$ such that for any $\beta \in (0, 1]$, we have that 
	\begin{itemize}
		\item{\textbf{$\Yes$ Case}: \[ \Pr_{\Dist \sim Q_{\Yes},  \TesterFunc}\left[ \TesterFunc^{(\Dist)}(\DomainSize, \beta) = \Yes\right] \geq 1-\beta\]}

	    \item{\textbf{$\No$ Case}: 
		     \[
			   \Pr_{\Dist \sim Q_{\No},  \TesterFunc}\left[ \TesterFunc^{(\Dist)}(\DomainSize, \beta) = \No\right] \geq 1-\beta\] } 
	\end{itemize}
\end{definition}

The main part of the argument will be to define a ``hard promise problem.'' At the end of this section, we will argue that this also gives us a lower bound for a specific fixed label-invariant property.

Let $K$ be a sufficiently large constant. The hard promise problem we consider will be the following:
\begin{align}\label{eq:hardprop}
		\Property_\Yes & \Def  \{ \Dist \in \DistSet{\Domain}: \SupportSize{\Dist} = \DomainSize^{2/3} / K \land \Dist \text{ is flat}\} \\
		\Property_\No  & \Def  \{ \Dist \in \DistSet{\Domain}: \SupportSize{\Dist} = \DomainSize^{2/3} \land \Dist \text{ is flat}\}
	\end{align}

We will let the meta-distributions $Q_{\Yes}$ and $Q_{\No}$ be the uniform distributions over each of the above properties.

We now state a theorem of Valiant that argues how one can use moment matching techniques to prove lower bounds for $\Samp$ testers for label-invariant properties. We state a modification of the theorem that applies for distinguishing between two label-invariant properties, and follows from the same proof as the original theorem. 
\begin{theorem}[Modified Version of Wishful Thinking Theorem \citep{valiantthesis}]\label{thm:wishful-thinking}
Fix label-invariant properties $\Property_{\Yes}, \Property_{\No}$. Given a positive integer $m$, a distance parameter $\Proximity$, and two distributions $p_{\Yes} \in \Property_{\Yes}, p_{\No} \in \Property_{\No}$, suppose the following conditions hold:
\begin{enumerate}
    \item No large elements:
    \[
    \| p_{\Yes} \|_{\infty}, \| p_{\No} \|_{\infty} \leq \frac{1}{500m},
    \]
    \item Moments (approximately) match:
    \[
    \sum_{j=2}^{\infty} \frac{|m^{\Yes}(j) - m^{\No}(j)|}{\lfloor j/2  \rfloor! \sqrt{1+ \max\{m^{\Yes}(j), m^{\No}(j)\}}} \leq \frac{1}{24},
    \]
    where $m^{\Yes}(j) = m^j \|p_{\Yes}\|_j^j$, $m^{\No}(j) = m^j \|p_{\No}\|_j^j$. 
\end{enumerate}
Then, any BPP tester for property promise $\Property_{\Yes}, \Property_{\No}$, with failure probability $\frac{1}{3}$ needs at least $m$ samples.
\end{theorem}

We now connect testing for meta-distributions with testing for property promise problems.

\begin{lemma}\label{lem:worstcasetoavgcase}
    Let $(\Property_{\Yes},    \Property_{\No}) \subseteq \Delta([N])$ be the properties defined in Equation~\ref{eq:hardprop}, and let $Q_{\Yes}$ and $Q_{\No}$ be the uniform distribution over $\Property_{\Yes}, \Property_{\No}$ respectively. If there exists a $\BPPTester$ tester for $Q_{\Yes}, Q_{\No}$ with failure probability $\beta$ and $\Samp$ access to the distribution that uses $\ell$ samples from the distribution, then there exists a $\BPPTester$ tester for property promise $\Property^{\Yes}, \Property^{\No}$ with failure probability $\beta$ and $\Samp$ access to the distribution that uses $\ell$ samples from the distribution.
\end{lemma}
\begin{proof}
    We reduce from distinguishing $\Property_{\Yes}$ and $\Property_{\No}$ in the worst case to distinguishing meta-distributions $Q_{\Yes}$ and $Q_{\No}$. Let $A$ be the algorithm that distinguishes $Q_{\Yes}$ and $Q_{\No}$ with probability at least $1-\beta$. 

    Fix any distribution $P \in \Property_{\Yes} \cup \Property_{\No}$. 
    Consider a random permutation $\pi: [N] \to [N]$ of the domain. Note that $\pi(P)$ is distributed according to $Q_{\Yes}$ (if $P \in \Property_{\Yes}$) or $Q_{\No}$ (if $P \in \Property_{\Yes}$). 
    
    Let $S \samples P^{\otimes \ell}$, and consider an algorithm $A'$ (with blackbox access to $A$) that first applies $\pi$ to each element of $S$. Observe that this process is equal to sampling the distribution from either $Q_{\Yes}$ or $Q_{\No}$ and then taking $\ell$ samples from the resulting distribution. 
    Now, if Algorithm $A'$ uses tester $A$ on the samples $\pi(S)$, tester $A$ can tell with probability at least $1-\beta$ over the randomness of the meta-distribution, the randomness of the samples $S$, and the internal randomness of the algorithm whether the distribution was $Q_{\Yes}$ or $Q_{\No}$. 
    Hence, $A'$ can distinguish with probability at least $1-\beta$ over its internal randomness (which includes the random permutation), and the randomness of the sample $S$ whether $P$ was in $\Property_{\Yes}$ or $\Property_{\No}$. 
\end{proof}

Next, we use the wishful thinking theorem to prove a lower bound on the hard property promise problem above in the $\Samp$ model. 

\begin{lemma}\label{lem:worstcasesamp}
   Fix $\beta \leq \frac{1}{3}$.  Any $\BPPTester$ tester for property promise $\Property_{\Yes}, \Property_{\No}$ (the ``hard promise problem'' described at the start of the section) with failure probability $\beta$ and $\Samp$ access to the distribution needs $\Omega(N^{1/3})$ samples. 
\end{lemma}
\begin{proof}
    We will leverage the ``wishful thinking theorem'' (Theorem~\ref{thm:wishful-thinking}) in order to prove this lower bound.

    We will choose $p_{\Yes} \in \Property_{\Yes}$ and $p_{\No} \in \Property_{\No}$ to be two distributions with maximally overlapping support, that is the support of the former is a subset of the latter. Let $m = C\cdot  N^{1/3}$ for a sufficiently small constant $C>0$. 

    We will verify the conditions of the theorem one at a time:

    \begin{enumerate}
        \item Firstly, $\|p_{\Yes}\|_{\infty} = \frac{K}{N^{2/3}} \leq \frac{1}{500 C N^{1/3}} = \frac{1}{500m}$, $\|p_{\No}\|_{\infty} = \frac{1}{N^{2/3}} \leq \frac{1}{500 C N^{1/3}} = \frac{1}{500m}$, where both inequalities hold for sufficiently large $N$. 
        \item Finally, we verify the condition on moments. Firstly, we compute $m^{\Yes}(j) = m^j  \| p_{\Yes} \|_j^j = \frac{m^j K^{j-1}}{N^{2/3(j-1)}}$, and $m^{\No}(j) = m^j  \| p_{\No} \|_j^j = \frac{m^j}{N^{2/3(j-1)}}$. 
        \begin{align*}
            \sum_{j=2}^{\infty} \frac{|m^{\Yes}(j) - m^{\No}(j) | }{\lfloor j/2 \rfloor! \sqrt{1+\max(m^{\Yes}(j) , m^{\No}(j))}} 
            &\leq  \sum_{j=2}^{\infty} |m^{\Yes}(j) - m^{\No}(j) | \\
            & 
            = \sum_{j=2}^{\infty} \frac{m^j K^{j-1}}{N^{2/3(j-1)} } \left( 1 - \frac{1}{K^{j-1}} \right) \\
            &\leq \sum_{j=2}^{\infty} \frac{m^j K^{j-1}}{N^{2/3(j-1)}} 
             \leq \sum_{j=2}^{\infty} \frac{C^j K^{j-1} N^{j/3}}{N^{2/3(j-1)}} \leq \sum_{j=2}^{\infty} C^j K^{j-1} \leq \frac{C^2 K}{1-C K} \leq \frac{1}{24},
        \end{align*}
    where the first inequality follows by reducing the denominator to $1$, the third inequality follows by substituting for $m$, the fourth inequality follows because $N^{j/3} \leq N^{2/3(j-1)}$ for $j \geq 2$, the fifth inequality follows by summing the geometric series and using the fact that $CK<1$ for sufficiently small $C$, and the last inequality follows since $C$ is a sufficiently small constant. 
    \end{enumerate}

    Then, invoking Theorem~\ref{thm:wishful-thinking}, we get that distinguishing between $\Property_{\Yes}$ and $\Property_{\No}$ requires at least $m$ samples, thereby proving the lower bound.
\end{proof}

Finally, we will argue that for this hard promise problem, we can simulate $\PCond$ queries, thereby proving a lower bound for testers with access to a $\PCond$ oracle. Intuitively, the main idea is that since the distributions in both $\Property_{\Yes}$ and $\Property_{\No}$ are flat and the supports are small, $\PCond$ queries are not useful: they only provide information when the query set $\{x,y\}$ include both one element outside the support and one inside, and if the number of samples is small then finding such a $\{x,y\}$ can only happen with vanishing probability.

\begin{theorem}[Hard Problem For $\PCond$ Tester]\label{thm:hard-label-invariant-pcond}

	Every BPP tester using a $\PCond$ oracle for property promise $\Property_{\Yes}$ and $\Property_{\No}$ with failure probability $0.01$ must make $\BigOmega{\DomainSize^{1/3}}$ queries.
	
\end{theorem}
\begin{proof}
We give a simulation argument to show how $\PCond$ queries can be simulated without using a $\PCond$ oracle.
Consider the following meta-distributions: $Q_{\Yes}$ is the uniform distribution over $\Property_\Yes$ and $Q_{\No}$ the uniform distribution over $\Property_\No$. Let $\Dist$ be the distribution drawn from the meta-distribution. The simulation argument below will be agnostic to which meta-distribution it is drawn from.


\noindent Consider the following oracle $\PCond^{\rm{}Sim}$ that simulates oracle $\PCond$ given access to a sample set $S$.

\begin{mdframed}[
		frametitle={$\PCond^{\rm{}Sim}$},
		frametitlealignment=\centering
	]
	\flushleft
	\textbf{Input}: Description of domain $\Domain$, Sample (Multi-set) $S \subseteq [N]$, query points $z_1, z_2$.\\

    \textbf{Output}: $x \in \{z_1, z_2, \bot\}$.\\
    \begin{itemize}
        \item If $z_1, z_2 \not\in S$, output $\bot$.
    
        \item If $z_1, z_2 \in S$, with probability $1/2$ output $z_1$ and with probability $1/2$ output $z_2$.

        \item If $z_1 \in S$, $z_2 \not\in S$ output $z_1$, and if  $z_1 \not\in S$, $z_2 \in S$, output $z_2$.
    \end{itemize}

\captionsetup{hypcap=false}
\captionof{figure}{Simulation of $\PCond$ Oracle with only access to samples from $\Dist$}
\label{alg:pcondsim}
\captionsetup{hypcap=true} 
\end{mdframed}

Any algorithm $A$ that uses a $\PCond$ oracle can be thought of without loss of generality in the following way: it takes $S$ consisting of $k$ i.i.d.\ samples from the distribution upfront, and then adaptively makes $\PCond$ queries $q_1 = (y_1, y_1'),\dots,q_m = (y_m, y_m')$ (with some computations in between).

For reasonable values of $m$ and $k$, we will now argue that any such algorithm can be simulated with the $\PCond^{\rm{}Sim}$ oracle, where the input $S$ is the sample drawn by algorithm $A$. In particular, we will consider $m,k \leq C' N^{1/3}$ for some constant $C'$.

Fix any sample $S$ for the rest of the argument. Define the event $E_r$ as follows: 
\[
E_r = \bigcap_{i=1}^r \left(y_i \in S \lor y_i \not\in \Support{\Dist} \right) \land  \left(y'_i \in S \lor y'_i \not\in \Support{\Dist} \right)\,.
\]

We first note that if event $E_r$ happens, then the output of $\PCond^{\rm{}Sim}$ when run on $[N], S, y_i, y_i'$ is identically distributed to the output of $\PCond$ when run on $[N], S, y_i, y_i'$ for all $i \in [r]$. To see this, we consider the different cases: when $y_i, y_i' \not\in S$, then by the definition of event $E_r$, $y_i, y_i' \not\in \Support{\Dist}$, and so the output of $\PCond$ would be $\bot$ (identical to the output of $\PCond^{\rm{}Sim})$. If  $y_i, y_i' \in S$, then by the definition of the promise problem, we have that $\Dist$ is uniform over its support and so the output of $\PCond$ is equiprobable between $y_i$ and $y_i'$, identical to the distribution of the output of $\PCond^{\rm{}Sim}$. And finally, if one of $y_i, y_i'$ is in  $S$ and the other is not in $S$ (say WLOG that $y_i \in S$), then by the definition of $E_r$, $y_i' \not\in \Support{\Dist}$, and so the output of $\PCond$ would be $y_i$, identical to the output of $\PCond^{\rm{}Sim}$.

Next, we bound the probability of $\overline{E_r}$ (to argue that this bad event happens with low probability).
\begin{align}\label{eq:eventErp}
    \Pr(\overline{E_r}) \leq \Pr(\overline{E_r} \mid E_1,\dots,E_{r-1}) + \sum_{i=1}^{r-1} \Pr(\overline{E_i})
\end{align}
    Consider the first term in the RHS above. Conditioning on $E_1,\dots,E_{r-1}$ occurring, we have that for $\overline{E_r}$ to occur, we need that either $y_r$ or $y_r'$ belong to $\Support{\Dist}$ but do not belong to sample $S$. Let $S' = S \cup \{y_1,\dots,y_{r-1}, y'_1,\dots,y'_{r-1} \}$ ($S'$ is defined as a set and hence does not have repeats). Note that we always have $|S'|\leq N/2$ (all sets that can occur satisfy this condition since $m,k \leq C' N^{1/3}$). Observe that since we conditioned on $E_1,\dots,E_{r-1}$, each element of $S'$ either belongs to $S$ or is not in $\Support{\Dist}$. Next, by symmetry, observe that all elements in $[N] \setminus S'$ have equal probability of being in the support of $\Dist$ conditioned on  $E_1,\dots,E_{r-1}$, and since $N-|S'| \geq N/2$ that probability is upper bounded by $\frac{\SupportSize{\Dist}- |S|}{N/2} \leq \frac{\SupportSize{\Dist}}{N/2} \leq \frac{N^{2/3}}{N/2} = \frac{2}{N^{1/3}}$. By a union bound, we get that $\Pr(\overline{E_r} \mid E_1,\dots,E_{r-1}) \leq \frac{4}{N^{1/3}}$. 

Hence, substituting back in Equation~\ref{eq:eventErp}, we get that $\Pr(\overline{E_m}) \leq \frac{4m}{N^{1/3}}$. Thus, with probability at least $1-\frac{4m}{N^{1/3}}$ over the choice of the distribution and the samples from the distribution, we get that Algorithm $A$ run with $\PCond^{\rm{}Sim}$ perfectly simulates Algorithm $A$ run with $\PCond$. Also observe that $A$ run with $\PCond^{\rm{}Sim}$ is a BPP tester that makes only $\Samp$ queries. 

Now, for an appropriately chosen constant $C'>0$, assume by way of contradiction that algorithm $A$ making $C' N^{1/3}$ queries to the $\PCond$ oracle is a BPP tester for property promise $\Property_{\Yes}$ and $\Property_{\No}$ with failure probability at most $0.01$. This implies that algorithm $A$ making $C' N^{1/3}$ queries to the $\PCond$ oracle is a BPP tester for $Q_{\Yes}$ and $Q_{\No}$ with failure probability at most $0.01$. By the above argument $A$ making $C' N^{1/3}$ queries to the $\PCond^{\rm{}Sim}$ oracle is a BPP tester for $Q_{\Yes}$ and $Q_{\No}$ with failure probability at most $0.01 + \frac{4m}{N^{1/3}} \leq 0.02$ where the last inequality holds for sufficiently small $C'$ using the fact that $m \leq C' N^{1/3}$. This guarantees us that there is a BPP tester for $Q_{\Yes}$ and $Q_{\No}$ that makes only $\Samp$ queries with failure probability at most $0.02$ that uses $C' N^{1/3}$ samples from the distribution.

However, by Lemmas~\ref{lem:worstcasetoavgcase} and~\ref{lem:worstcasesamp}, we get that any BPP tester with $\Samp$ access to the distribution that distinguishes $Q_{\Yes}$ and $Q_{\No}$ with probability at least $0.98$ over the draw of the distribution, the samples from the distribution, and the internal randomness of the algorithm needs $\Omega(N^{1/3})$ samples. This gives a contradiction, thereby proving the theorem.\qedhere



\end{proof}

Finally, observe that $d_{TV}(\Property_{\Yes}, \Property_{\No} ) > \frac{1}{2}$ for sufficiently large support sizes, which also means we have proved the following result.

\begin{corollary}\label{cor:lowerbound}
    Every $\BPPTester$ tester using a $\PCond$ oracle for property $\Property_{\Yes}$ with proximity parameter  $\Proximity \leq 1/2$ and failure probability  $0.01$ must make $\BigOmega{\DomainSize^{1/3}}$ queries.
\end{corollary}

\begin{remark}
    We note that with the techniques above, the lower bound of $\Omega(\DomainSize^{1/3})$ is the best we can hope to prove by varying the support sizes in the promise problems. In choosing these support sizes $s_1 < s_2$ in defining $\Property_{\Yes}, \Property_{\No}$, we trade-off between two parameter settings, the error probability of the simulation oracle (where we simulate $\PCond$ queries using $\Samp$ in the proof of Theorem~\ref{thm:hard-label-invariant-pcond}), which is $O(m s_2 /\DomainSize)$, where $m$ is the number of $\PCond$ queries made, and the lower bound obtained for the $\Samp$ model, which would be $m = \Omega(\sqrt{s_1})$. Balancing these inequalities (and also noting that the error probability needs to be sufficiently small) gives that setting $s_1, s_2 \approx N^{2/3}$ is optimal, motivating our choices and giving a lower bound of $\Omega(N^{1/3})$. It is an open question whether this lower bound can be improved by tweaking our approach or by other techniques.
\end{remark}

\section{Support Size Verification}\label{sec:suppsize}

In this section, we describe a succinct proof system for the two-sided support size decision problem, which may be of independent interest. 
This proof system, when used in conjunction with the \nameref{lemma:estimate-neighborhood}, allows a tester to accurately approximate the mass of ``important`` points in \emph{any} ``relevant/heavy'' bucket of the $(\DomainSize, \Proximity)$-histogram for the distribution.
This ability to estimate the mass of points in heavy enough buckets is critical for verifying the prover's claimed histogram, which then allows us to verify any label-invariant property.
Our key technical contribution here is to ensure that the communication complexity of the proof system is sub-linear in the domain size $N$ while also ensuring polylogarithmic in $N$ query and sample complexities.
To ensure low communication, we give two different protocols to apply based on the parameter range of interest.

\subsection{Proofs For Testing Support Size For Distributions With Large Support}

To start with, we describe with Figure \ref{alg:large-supp-size}, a proof system for distributions with large support, which requires us to run two separate proof systems sequentially for the two sides of the test (testing whether the claimed support is too large or too small).

\begin{mdframed}[
		frametitle={Uniform Protocol},
		frametitlealignment=\centering
	]
	\flushleft
	\textbf{Common Input}: Description of domain $\Domain$, parameters $\alpha,\delta\in(0,1]$, and $A', A, B, B'$ with $\alpha \leq \frac{3}{2}\frac{B'-B}{B}$.\\

	\textbf{Verifier Input}: $\PCond$ access to an $(1+\alpha)$-flat distribution $\Dist \in \DistSet{\Domain}$.\\

	\textbf{Prover Input}: Full description of $\Dist$.\newline

	Draw $\highlight{x} \samples \Dist$. \newline

	\textbf{Test 1:} ``\textsc{Detect Small Support}''
    \begin{itemize}
        \item Compute the values $p_{\Reject},$ $q_{\Accept}$ as in~\cref{claim:test1}, and set
        \[
            \Delta_1 = (1-q_{\Accept}) - p_{\Reject} = \BigOmega{\left(\frac{A-A'}{A}\right)^2}
        \]
        \item Run sequentially the proof system \textbf{Test~1} of~\cref{alg:large-supp-size:test1}, on input $\True{x}$, $A,A'$ a total of
        \[
            T_1 \Def \BigO{\frac{\log(1/\delta)}{\Delta_1^2}}
        \]
        times, with fresh randomness for each of the $T_1$ runs. Let $0\leq t_1\leq T_1$ be the number of runs in which the proof system returns $\Accept$. 
        \item Set the outcome of this test to $\Accept$ if $t_1 < (p_{\Reject} + \frac{\Delta_1}{2})T_1$, and to $\Reject$ otherwise.
    \end{itemize}

	\textbf{Test 2:}  ``\textsc{Detect Large Support}''
    \begin{itemize}
        \item Compute the values $p'_{\Reject},$ $q'_{\Accept}$ as in~\cref{claim:test2}, and set
        \[
            \Delta_2 = (1-q'_{\Accept}) - p'_{\Reject} = \BigOmega{\frac{\left(B'-B\right)^2}{B'^2}\frac{B}{B'}}
        \]
        \item Run sequentially the proof system \textbf{Test~2} of~\cref{alg:large-supp-size:test2}, on input $B,B'$ a total of
        \[
            T_2 \Def \BigO{\frac{\log(1/\delta)}{\Delta_2^2}}
        \]
        times, with fresh randomness for each of the $T_2$ runs. Let $0\leq t_2\leq T_2$ be the number of runs in which the proof system returns $\Accept$. 
        \item Set the outcome of this test to $\Accept$ if $t_2 < (p'_{\Reject} + \frac{\Delta_2}{2})T_2$, and to $\Reject$ otherwise.
    \end{itemize}
    
	\textbf{Final Output}: The verifier outputs $\Accept$ if and only if both tests above were set to $\Accept$.

\captionsetup{hypcap=false}
\captionof{figure}{Proof System For Support Size Range For Distributions With Large Support}
\label{alg:large-supp-size}
\captionsetup{hypcap=true} 
\end{mdframed}

\begin{lemma}[Support Size Difference For Large Support Distributions]\label{lemma:supp-size-large}Fix $\DomainSize \in \Naturals$, and let $0\leq \alpha \leq 1$.
	Fix integer parameters $A' < A < B < B'$ such that $A \ge \sqrt{\DomainSize}$ and $\alpha \leq \frac{3}{2}\cdot \frac{B'-B}{B}$, and failure probability $\delta\in(0,1]$.
	Given $\PCond$ access to a $(1+\alpha)$-flat distribution $\Dist \in \DistSet{\Domain}$, the interactive proof system described in Figure~\ref{alg:large-supp-size}  decides the following promise problem completeness and soundness errors are at most $\delta$:
    \begin{align}
		\Property_\Yes & \Def  \{ \Dist \in \DistSet{\Domain}: A \le \SupportSize{\Dist} \le B \land \Dist \text{ is $(1+\alpha)$-flat}\}                    \\
		\Property_\No  & \Def  \{ \Dist \in \DistSet{\Domain} : \left(\SupportSize{\Dist} \le A' \lor \SupportSize{\Dist} \ge B'\right) \land \Dist \text{ is $(1+\alpha)$-flat}\}
	\end{align}
    The proof system has communication complexity
    \[
    \BigO{\max\left(
        \frac{\DomainSize}{A}\cdot \left(\frac{A}{A-A'}\right)^4,
        \frac{\DomainSize}{B'}\cdot \left(\frac{B'}{B}\right)^2\left(\frac{B'}{B'-B}\right)^4  \log \frac{1}{\delta}\right)}\,
    \]
    sample complexity
    \[
        \BigO{\left(\frac{B'}{B}\right)^2\left(\frac{B'}{B'-B}\right)^4 \log \frac{1}{\delta}}\,,
    \]
    query complexity
    \[
        \BigOTilde{\left(\frac{A}{A-A'}\right)^4\log \frac{1}{\delta}}\,,
    \]
	and round complexity
    \[
    \BigO{\max\left(
        \left(\frac{A}{A-A'}\right)^4,
        \left(\frac{B'}{B}\right)^2\left(\frac{B'}{B'-B}\right)^4 
    \right) \log \frac{1}{\delta}}\,.
    \]
\end{lemma}
\begin{proof}
The various complexities immediately follow from those of~\cref{claim:test1,claim:test2}, after repetition for $T_1 = \BigO{
        \left(\frac{A}{A-A'}\right)^4 \log \frac{1}{\delta}}$ and $T_2 = \BigO{
        \left(\frac{B'}{B}\right)^2\left(\frac{B'}{B'-B}\right)^4 \log \frac{1}{\delta}}$ rounds, respectively.
The stated completeness and soundness errors follow from (sequential) repetition with a threshold rule, along with a standard Hoeffding bound.\footnote{Namely, if a single test has rejection probabilities $p,q$ with $q\geq p+\Delta$ in the $\Yes$ and $\No$ cases, respectively, repeating the test sequentially (with fresh randomness) a total of $O(\log(1/\delta)/\Delta^2)$ times and thresholding at $p+\frac{\Delta}{2}$ results in a correct outcome with probability at least $1-\frac{\delta}{2}$.} Amplifying both \textbf{Test~1} and \textbf{Test~2} by sequential repetition to correctly estimate their rejection probability, except with probability $\frac{\delta}{2}$, and taking a union bound over the two, yields the claimed statement.
\end{proof}
\begin{mdframed}[
		frametitle={Test 1},
		frametitlealignment=\centering
	]
	\flushleft
	\textbf{Common Input}: Description of domain $\Domain$, integers parameters $A' < A$.\\

	\textbf{Verifier Input}: $\PCond$ access to a nearly flat distribution $\Dist \in \DistSet{\Domain}$, element $\True{x} \in \Support{\Dist}$.\\

	\textbf{Prover Input}: Full description of $\Dist$.

	\begin{enumerate}
		\item Set $c\Def \frac{A-A'}{2A}$, $s \Def c\frac{\DomainSize}{A}$, $\beta \Def c^2 e^{-c}$.
		\item{\textbf{$\TesterFunc \rightarrow \Prover{\Dist}$}: Send set $Y$ where $Y = (y_1, \ldots, y_{s}) \samples \Uniform{\DomainSize}$, and ask the prover to send back any $\widetilde{y}\in Y$ such that $\widetilde{y} \in \Support{\Dist}$.         }

		\item{\textbf{$\TesterFunc \leftarrow \Prover{\Dist}$}: Sends back $\widetilde{y} \in \Support{\Dist}$. If no such $\widetilde{y}$ exists, then output $\Fail$.
		      }

		\item{$\TesterFunc$ computation: If prover outputs $\Fail$, then output $\Reject$. Otherwise, check $\widetilde{y} \in \Support{\Dist}$ using the \nameref{alg:pcond-support-checker}  with reference $\highlight{x}$ and probability of failure $\beta$: if it returns $\Yes$, output $\Accept$, and $\Reject$ otherwise.}

	\end{enumerate}
    \captionsetup{hypcap=false} 
	\captionof{figure}{First Proof System For Support Size Range For Distributions With Large Support}\label{alg:large-supp-size:test1}. 
    \captionsetup{hypcap=true} 
\end{mdframed}
\begin{claim}[Detecting Small Support]
    \label{claim:test1}
    The proof system given in~\cref{alg:large-supp-size:test1} has the following guarantees. 
    On input access to a nearly flat distribution $\Dist$, along with an element $\True{x}\in\Support{\Dist}$ and integers $A'< A \leq \DomainSize$, it has communication complexity $\BigO{\DomainSize/A}$, query complexity $\BigO{\log \frac{A}{A-A'}}$ (and sample complexity $0$), and  decides between
    (1)~$\SupportSize{\Dist} \geq A$ and (2)~$\SupportSize{\Dist} \leq A'$
    with completeness and soundness errors at most $p_{\Reject}$ and $q_{\Accept}$, explicit values such that
    \[
        1-q_{\Accept} > p_{\Reject} + \BigOmega{\left(\frac{A-A'}{A}\right)^2}\,.
    \]
    Moreover, it has round complexity $1$.
\end{claim}
\begin{proof}
    The communication complexity follows from our choice of $s = \BigO{\DomainSize/A}$, and the query complexity from the call to \nameref{alg:pcond-support-checker}, which has query complexity
    \[
        \BigO{\log\frac{1}{\beta}} = \BigO{\log\frac{1}{c}+c} = \BigO{\log\frac{1}{c}}= \BigO{\log\frac{A}{A-A'}}
    \]
    by our setting of parameters and~\cref{lemma:isinsupport}. Turning to correctness, we analyze the completeness and soundness claims separately.
    \begin{description}
        \item[Completeness:] assume that $\SupportSize{\Dist} \geq A$. We have the probability that a uniformly sampled point is in the support is at least $\frac{A}{\DomainSize}$.
	    If any of the samples $y_1, \ldots, y_{s}$ is in the support of $\Dist$, then the honest prover is able to get the verifier to accept in \textbf{Test 1} by sending one of these samples, unless the call to~\nameref{alg:pcond-support-checker} then fails (which since the distribution is nearly flat happens with probability at most $\beta$ by~\cref{lemma:isinsupport}).
	    Therefore, by a union bound, the completeness error for the first test is upper bounded as follows:
        \begin{align*}
    		&\Prob{\TesterFunc \text{ outputs } \Reject } \\
            &=  \PProb{ \{y_1, \ldots, y_{s}\}\cap \Support{\Dist} = \emptyset }{y_1, \ldots, y_{s} \samples  \Domain} 
            \\&\qquad + \PProb{ \{y_1, \ldots, y_{s}\}\cap \Support{\Dist} \neq \emptyset, \text{ \nameref{alg:pcond-support-checker} incorrect}}{y_1, \ldots, y_{s} \samples  \Domain} \\
            &\leq  \left(1 - \frac{A}{\DomainSize}\right)^{s}  + \beta\\
            &\leq  \left(1 - \frac{A}{\DomainSize}\right)^{c\frac{\DomainSize}{A}}  + \beta
            \\
            &\leq  e^{-c}  + c^2 e^{-c} = (1+c^2)e^{-c} = p_{\Reject}\,.
    	\end{align*} 
        
        recalling our choice of $s,\beta$, and using the inequality $1-x\leq e^{-x}$.
        \item[Soundness:] conversely, assume that $\SupportSize{\Dist} \leq A'$. \textbf{Test 1} can only accept if at least one of the uniformly chosen $y_1, \ldots, y_{s}$ falls in the support of $\Dist$ (otherwise it will always reject, as the~\nameref{alg:pcond-support-checker} has one-sided error in that case). By a union bound over all $s$ uniform samples, this leads to
        	\begin{align*}
        		\Prob{\TesterFunc \text{ outputs } \Accept \text{ in  \textbf{Test 1} }} 
                &\le  \PProb{\exists j \in [s] \text{ s.t. } y_j \in \Support{\Dist} }{y_1, \ldots, y_{s_1} \samples  \Uniform{\DomainSize}} \\
        		  & \le \frac{A'}{\DomainSize}s = q_{\Accept}\,.
        	\end{align*}
        Equivalently, multiplying and dividing $q_{\Accept}$ by $A$, and using the fact that $1-2c = \frac{A'}{A}$ from the setting of $c$,
        	\begin{align*}
        		\Prob{\TesterFunc \text{ outputs } \Reject} 
                &\geq  1 - \frac{A'}{A}\cdot \frac{A}{\DomainSize}s
                = 1 - c\cdot\frac{A'}{A}
                = 1 - c(1-2c) = 1-c+2c^2 = q_{\Reject}\,.
        	\end{align*}
    \end{description}
    Now, using the fact that $e^{-c} \leq 1-c+c^{2}/2$
    \[
        1-c+2c^2  - (1+c^2)e^{-c} \geq \frac{c^2}{2}
    \]
    
    which means the gap between the rejection probability in the completeness and soundness cases is at least
    \[
    q_{\Reject} - p_{\Reject} \geq \frac{c^2}{2} = \BigOmega{\left(\frac{A-A'}{A}\right)^2}
    \]
\end{proof}
\begin{mdframed}[
		frametitle={Test 2},
		frametitlealignment=\centering
	]
	\flushleft
	\textbf{Common Input}: Description of domain $\Domain$, integers parameters $B < B'$.\\

	\textbf{Verifier Input}: $\PCond$ access to an $(1+\alpha)$-flat distribution flat distribution $\Dist \in \DistSet{\Domain}$, where $\alpha \leq \frac{3}{2}\frac{B'-B}{B}$.\\

	\textbf{Prover Input}: Full description of $\Dist$.

	\begin{enumerate}
		\item Set $c\Def \frac{B'-B}{18 B'}$, $s \Def c\frac{\DomainSize}{B'}$.
        \item \textbf{$\TesterFunc$}: Draw $\True{x} \samples \Dist$ and $(z_1, \ldots, z_{s}) \samples \Uniform{\Domain}$. Form the tuple $S \Def (\highlight{x}, z_1, \ldots, z_{s})$.
		\item{\textbf{$\TesterFunc \rightarrow \Prover{\Dist}$}: Send the permuted tuple $S' \Def \{S_{\pi(1)}\dots, S_{\pi(s+1)}\}$, where $\pi$ is a permutation of $[s+1]$ chosen uniformly at random.}

		\item{\textbf{$\TesterFunc \leftarrow \Prover{\Dist}$}: Return $\widetilde{y}\samples \{i \in [s+1]: S'_{i} \in \Support{\Dist}\}$, an index chosen uniformly at random among all those for which the corresponding element of $S'$ is in the support of $\Dist$.}

		\item{If $\widetilde{y}\neq \pi(1)$ (i.e., $\widetilde{y}$ does not correspond to the index of $\True{x}$) output $\Reject$, otherwise output $\Accept$.}

	\end{enumerate}
    \captionsetup{hypcap=false} 
	\captionof{figure}{Second Proof System For Support Size Range For Distributions With Large Support}\label{alg:large-supp-size:test2}
    \captionsetup{hypcap=true} 
\end{mdframed}
\begin{claim}[Detecting Large Support]
    \label{claim:test2}
    The proof system given in~\cref{alg:large-supp-size:test2} has the following guarantees. 
    On input access to an $(1+\alpha)$-flat distribution $\Dist$, along with an element $\True{x}\in\Support{\Dist}$ and integers $B< B' \leq \DomainSize$ such that $0\leq \alpha \leq \frac{3}{2}\frac{B'-B}{B}$, it has communication complexity $\BigO{\DomainSize/B'}$, sample complexity $1$ (and query complexity $0$), and  decides between
    (1)~$\SupportSize{\Dist} \leq B$ and (2)~$\SupportSize{\Dist} \geq B'$
    with completeness and soundness errors at most $p'_{\Reject}$ and $q'_{\Accept}$, explicit values such that
    \[
        1-q'_{\Accept} > p'_{\Reject} + \BigOmega{\left(\frac{B'-B}{B'}\right)^2\frac{B}{B'}}\,.
    \]
    Moreover, it has round complexity $1$.
\end{claim}
\begin{proof}
    The communication complexity follows from our choice of $s = \BigO{\DomainSize/B'}$, and the sample complexity from the draw of $\True{x}$ (only call to the oracle for $\Dist$).
    
    Turning to correctness, we first define the following three disjoint events:
    \begin{itemize}
        \item $\mathcal{E}_0$: none of the $s$ uniformly drawn elements $z_1,\dots z_s$ falls in the support of the distribution; that is, $|\{z_1, \ldots, z_{s}\} \cap \Support{\Dist}| = 0$.

        \item $\mathcal{E}_1$: exactly one of the $s$ uniformly drawn elements $z_1,\dots z_s$ falls in the support of the distribution; that is, $|\{z_1, \ldots, z_{s}\} \cap \Support{\Dist}| = 1$.
        \item $\mathcal{E}_{\geq 2}$: at least two of the $s$ uniformly drawn elements $z_1,\dots z_s$ falls in the support of the distribution; that is, $|\{z_1, \ldots, z_{s}\} \cap \Support{\Dist}| \geq 2$.
    \end{itemize}
    The intuition is that a prover (honest or not) is able to send back the index of the verifier's true sample $\True{x}$ whenever $\mathcal{E}_0$ holds; under $\mathcal{E}_1$, while a prover cannot always send the index of $\True{x}$, there is still a strategy for them to guess it (among two alternatives) with probability $\approx 1/2$. Under $\mathcal{E}_{\geq 2}$, there is not much we can say, but fortunately this is a very unlikely event in the completeness case.
    
    With this in hand, we analyse the completeness and soundness claims. For convenience, set
    \[
    \lambda \Def  \frac{B}{B'} < 1
    \]
    and observe that our assumption on $\alpha$ implies $\alpha \leq \frac{3}{2}\cdot \frac{1-\lambda}{\lambda}$.
    \begin{description}
        \item[Completeness:] Assume $\SupportSize{\Dist} \leq B$. 
        As discussed above, if $\mathcal{E}_0$ holds then the honest prover succeeds, as it can uniquely identify $\True{x}$ in $S'$. If $\mathcal{E}_1$ holds (in which case $|S\cap \Support{\Dist}|=2$), then the honest prover can still succeed in returning the index of $\True{x}$ with probability $1/2$, by returning one of the two possible indices uniformly at random. If $\mathcal{E}_{\geq 2}$ holds, 
        while there is still a positive probability of guessing the index of $x$ correctly, we can (conservatively) neglect it for the analysis and count $\mathcal{E}_{\geq 2}$ as a failure. This leads to the following bound for the rejection probability, where we use the fact that
        $|\{z_1, \ldots, z_{s}\} \cap \Support{\Dist}|$ is a Binomial random variable with parameters $s$ and $\frac{|\Support{\Dist}|}{\DomainSize} \leq \frac{B}{\DomainSize}$:
		      \begin{align*}
			      \Prob{\TesterFunc \text{ outputs } \Reject } 
                  &\leq  0\cdot \Prob{\mathcal{E}_0} + \frac{1}{2}\cdot\Prob{\mathcal{E}_1} + 1\cdot \Prob{\mathcal{E}_{\geq 2}}   \\
                  &\leq  \frac{1}{2}\cdot s\frac{B}{\DomainSize}\left(1-\frac{B}{\DomainSize}\right)^{s-1} + \Prob{\mathcal{E}_{\geq 2}}   \\
                  &=  \frac{c\lambda}{2}\left(1-\frac{c\lambda}{s}\right)^{s-1} + \Prob{\mathcal{E}_{\geq 2}}   \\
                  &\leq  \frac{c\lambda}{2}\frac{e^{-c\lambda}}{1-\frac{c\lambda}{s}} + \binom{s}{2}\left(\frac{B}{N}\right)^2   \\
                  &\leq  \frac{c\lambda}{2}\frac{1}{1-\frac{c\lambda}{2}} + \frac{\left(c\lambda\right)^2}{2} \\
                  &\leq  \frac{c\lambda}{2} + \left(c\lambda\right)^2
                  \leq  \frac{c\lambda}{2} + c^2\lambda = p'_{\Reject}\,,
		      \end{align*}
		      by our setting of $s= c\frac{\DomainSize}{B'}$, and crudely bounding
              $\Prob{\mathcal{E}_{\geq 2}} \leq \binom{s}{2}\left(\frac{B}{N}\right)^2 \leq \frac{s^2}{2} \left(\frac{c\lambda}{s}\right)^2$ before concluding with $\frac{1}{1-x} \leq 1+2x$ (which holds for $x\in[0,1/2]$).
            Importantly, we here implicitly used the fact that $\Pr[\mathcal{E}_1] = \PProb{X=1}{X\sim \operatorname{Bin}(s,p)} = sp(1-p)^{s-1}$ is non-decreasing in the Binomial parameter $p$, as $\frac{B}{\DomainSize}$ is only an upper bound on this parameter. One can check that this is true as long as $p \leq \frac{1}{s+1}$, which is satisfied for us, as, given our setting of $s$ and $c$, we have 
            \[
                p \leq \frac{B}{\DomainSize} \leq \frac{B}{B'}
                \leq 9\frac{\DomainSize}{B'-B}
                = \frac{1}{2s} \leq 
                \frac{1}{s+1}.
            \]
   
        \begin{remark}
          Note that the above analysis critically saves (almost) a factor $2$ over a naive union bound, which using the above notation would have given $\leq c\lambda$. 
          Instead, we get (almost) $\leq \frac{c\lambda}{2}$. 
          This is critical, as for sequential repetition, we need $p'_{\Reject} < q'_{\Reject} \approx \frac{1}{2}(1-e^{-c})$, which would have been impossible if we didn't save this factor 2, given that $\lambda\approx 1-\Proximity$ can be very close to 1 in the settings we use these protocols in.
        \end{remark}
        \item[Soundness:] Assume $\SupportSize{\Dist} \geq B'$. If $|S\cap \Support{\Dist}|\geq 2$ (i.e., if $\bar{\mathcal{E}}_0 = \mathcal{E}_1\cup \mathcal{E}_{\geq 2}$ holds), we claim that the prover can only return the index of $\True{x}$ (and thus make the tester accept) with probability at most $\frac{1+\alpha}{2+\alpha} = \frac{1}{2} + O(\alpha)$. 
        A dishonest prover maximizes its chance of returning the index of $\True{x}$ by choosing the index with the highest conditional probability of being $\pi(1)$ given the permuted tuple $S'$. By Bayes' rule, this conditional probability of an index is proportional to the probability mass of the corresponding element under $\Dist$. Let $z_i$ be another element in $\Support{\Dist}$. 
        Then since the distribution is $(1+\alpha)$-flat, the dishonest prover chooses the correct index of $x$ with probability at most
            \[
                \frac{\Dist[x]}{\Dist[x]+\Dist[z_i]} \leq \frac{1+\alpha}{2+\alpha}.
            \]
        Of course, if $|S\cap \Support{\Dist}| = 1$ (i.e., if $\mathcal{E}_0$ holds) then the prover can always identify $\True{x}$ and return the correct index. Hence, we have
		      \begin{align*}
			      \Prob{\TesterFunc \text{ outputs } \Accept} & = \Prob{\text{Prover guesses $\True{x}$ from $S$}}\Prob{\bar{\mathcal{E}}_0}+ 1\cdot \Prob{\mathcal{E}_0}  \\
			      & \le \frac{1+\alpha}{2+\alpha}\Prob{\bar{\mathcal{E}}_0}+ \Prob{\mathcal{E}_0}   \\
			    & = \frac{1+\alpha}{2+\alpha} + \frac{1}{2+\alpha} \Prob{\mathcal{E}_0}      \\
			      & \le \frac{1+\alpha}{2+\alpha} + \frac{1}{2+\alpha}  \left(1 - \frac{B'}{\DomainSize}\right)^{s} \\
                & \leq \frac{1+\alpha}{2+\alpha} + \frac{1}{2+\alpha}  e^{-c} = q'_{\Accept},
		      \end{align*}
              the last inequality from our choice of $s= c\frac{\DomainSize}{B'}$. Equivalently,
        	\begin{align*}
        		\Prob{\TesterFunc \text{ outputs } \Reject} 
                &\geq  1 -  q'_{\Accept}
                = \frac{1}{2+\alpha}\left(1-e^{-c}\right) 
                \geq \frac{1}{2}\cdot\frac{1-e^{-c}}{1+\frac{3}{4}\cdot \frac{1-\lambda}{\lambda}} = q'_{\Reject}\, ,
        	\end{align*}
    \end{description}
    where the last inequality follows by the assumption that $\alpha \leq \frac{3}{2} \cdot \frac{B' - B}{B}$.
    Now, by our setting of $c = \frac{B'-B}{18 B'} = \frac{1-\lambda}{18}$, we have 
        \[
            4\frac{1-e^{-c}}{(1+4c)c} - 3 \geq 1-18c = \lambda
        \]
        where the first inequality follows from convexity of the function $f\colon x\mapsto 4\frac{1-e^{-x}}{(1+4x)x} - 3$ (extended at $0$ by continuity), above its tangent $x\mapsto 1-18x$ at $0$.  
        Reorganizing and recalling the expressions of $p'_{\Reject},q'_{\Reject}$, is equivalent to
        \[
            q'_{\Reject} \geq p'_{\Reject} + c^2\lambda\,,
        \]
        concluding the proof since $c^2\lambda = \BigTheta{\left(\frac{B'-B}{B'}\right)^2\frac{B}{B'}}$.
        
\end{proof}
\subsection{Proofs For Testing Support Size For Distributions With Small Support}

The techniques described above to failed to give us sub-linear communication when dealing with distributions with small supports. 
To deal with this case, we describe a non-interactive proof system with sublinear communication, that decides the support size decision problem.

\begin{mdframed}[
		frametitle={Small Support Size Protocol},
		frametitlealignment=\centering
	]
	\textbf{Common Input}: Description of domain $\Domain$, parameters $A', A, B, B'$ where $B \le \sqrt{\DomainSize}$. Soundness parameters $\delta_{s}$ and completeness parameters $\delta_c$.\\
	\textbf{Verifier Input}: $\PCond$ access to a $(1+\alpha)$-flat distribution $\Dist \in \DistSet{\Domain}$.\\
	\textbf{Prover Input}: Full description of $\Dist$.\\
	\flushleft
	Set $s_1 = \left\lceil\frac{\log(1/\delta_{s})}{\log(A/A')}\right\rceil$,  and     
    $s_2 = \left\lceil\frac{\log(1/\delta_{s})}{\log((1+\alpha)B'/B)}\right\rceil$
    \newline

	Prover sends a set $\Claimed{\Support{\Dist}}$ of size at most $B$ (and at least $A$), which it claims is the support of $\Dist$.
	If the set size is outside this range, the verifier immediately outputs $\Reject$.\newline

	\textbf{Test 1:} ``\textsc{Detect Small Support}''\newline

	Sample $z_1, \ldots, z_{s_1} \iidSamples \Claimed{\Support{\Dist}}$ uniformly, and $\True{x}\samples \Dist$. Check if $z_j \in \Support{\Dist}$, for all $j \in [s_1]$, using the \nameref{alg:pcond-support-checker} algorithm with error parameter $\frac{\delta_c}{s_1}$ and reference point $\True{x}$.
	Output $\Reject$ if \emph{any} invocation outputs reject.
	 \newline

	\textbf{Test 2:} ``\textsc{Detect Large Support}''\newline

	Verifier samples  $\highlight{x_1}, \ldots, \highlight{x_{s_2}} \iidSamples \Dist$ and outputs $\Reject$ Test 2 if any of the samples are not in $\Claimed{\Support{\Dist}}$.  \newline

	\textbf{Final Output}: The verifier outputs $\Accept$ if and only if it accepts both tests above.

\captionsetup{hypcap=false} 
	\captionof{figure}{Proof System For Support Size Range For Distributions With Small Support}\label{alg:small-supp-size}
    \captionsetup{hypcap=true} 
\end{mdframed}

\begin{lemma}[Support Size Difference For Small Support Distributions]\label{lemma:supp-size-small}
	Fix parameters for completeness error $\delta_c \in (0,1)$ and soundness error $\delta_s \in (0,1)$.
	Let $\DomainSize \in \Naturals$ be the domain size and $\alpha \in [0,1]$.
	Fix integer parameters $A' < A < B < B'$.
	Given $\PCond$ access to an $(1+\alpha)$-flat distribution $\Dist \in \DistSet{\Domain}$, the an interactive proof system described in Figure~\ref{alg:small-supp-size}  decides the following promise problem with communication complexity $\BigO{B}$.
	\begin{align}
		\Property_\Yes & \Def  \{ \Dist \in \DistSet{\Domain}: A \le \SupportSize{\Dist} \le B \land \Dist \text{ is  $(1+\alpha)$-flat}\}                   \\
		\Property_\No  & \Def  \{ \Dist  \in \DistSet{\Domain}: \left(\SupportSize{\Dist} \le A' \lor \SupportSize{\Dist} \ge B'\right) \land \Dist \text{ is $(1+\alpha)$-flat}\}
	\end{align}

The sample complexity is $\BigO{\frac{\log (1/\delta_s)}{\log(B'/B)}}$ and the query complexity is $\BigOTilde{\frac{\log(1/\delta_s)\log(1/\delta_c)}{\log(A/A')}}$.
\end{lemma}

\begin{proof}
The communication complexity is immediate from the cost of sending the purported support $\Claimed{\Support{\Dist}}$ (of size at most $B$). The sample and query complexities are, respectively, $O(s_2)$ and $O(s_1\log\frac{s_1}{\delta_c})$ (from the $s_1$ calls to the~\nameref{alg:pcond-support-checker} with error parameter $\delta_c/s_1$); and the claimed bounds follow from our choice of $s_1,s_2$ and (for the latter) the query complexity of~\cref{lemma:isinsupport}. We now turn to correctness, arguing completeness and soundness separately.\\

	\textbf{Completeness}: If the prover follows the protocol honestly, then Test 2 always passes.
	The probability that Test 1 fails is equal to the probability that the \nameref{alg:pcond-support-checker} fails, which, due to our assumption of near flatness and our choice of $s_1$, is at most $\delta_c$ by~\cref{lemma:isinsupport}, along with a union bound over $s_1$ tests.   \\


	\textbf{Soundness:} Suppose now that $\SupportSize{\Dist} \notin (A',B')$. We distinguish two cases, showing that \textbf{Test 1} will reject with probability at least $1-\delta_s$ if $\SupportSize{\Dist} \leq A'$, and \textbf{Test 2} will reject with probability at least $1-\delta_s$ if $\SupportSize{\Dist} \geq B'$.

    \begin{itemize}
        \item Suppose that $\SupportSize{\Dist} \leq A'$. We then have that, for each $1\leq i\leq s_1$, $\Pr[z_i \in \Support{\Dist}] \;\le\; \frac{A'}{A}$. 
	The tester accepts Test 1 if \emph{every} $z_i \in \Claimed{\Support{\Dist}}$ is found to be the support of $\Dist$.
	Thus, with $s_1$ independent samples,
	\begin{align*}
		\Prob{\TesterFunc \text{ outputs  }\Accept} 
        &= \Prob{\forall i,\; (z_i \in \Support{\Dist} \land \text{\nameref{alg:pcond-support-checker}  returns } \Yes)} \\
        &\leq
        \Prob{\forall i,\; z_i \in \Support{\Dist}}
        \leq 
		\left(\frac{A'}{A}\right)^{s_1}\,,
	\end{align*}
  Which is at most $\delta_s$ by our choice of $s_1 \ge \frac{\log (1/\delta_s)}{\log (A/A')}$.

	%

  \item Suppose now that $\SupportSize{\Dist} \geq B'$. 
    The tester accepts Test 2 if every $\highlight{x_1}, \ldots, \highlight{x_{s_2}}$ is in $\Claimed{\Support{\Dist}}$.
	Here we are checking if that cheating prover lied by claiming that $\Size{\Claimed{\Support{\Dist}}}$ is much smaller than the true support size.
	Intuitively, the best chance for the prover to get away with this is to use $ \Size{\Claimed{\Support{\Dist}}} = B$, when $\SupportSize{\Dist} = B'$ (as this is the smallest possible discrepancy between the cheating prover, and the correct answer), and then put the heaviest items of $\Dist$ in $\Claimed{\Support{\Dist}}$.
	This way, when the honest verifier samples $x \samples \Dist$, they are more likely to find that $x$ is in $\Claimed{\Support{\Dist}}$.
	The winning probability of this optimal cheating prover is maximised when the items in $\Claimed{\SupportSize{\Dist}}$ is heaviest.
	However, as the distribution is $(1+\alpha)$-flat, there is no such thing as an outright heavy or light element.
	This means that the provers strategy of declaring only $B$ elements as support is bound to miss out on items that are still highly likely to show up as samples. 

    More formally, observe first that since $\Dist$ is $(1+\alpha)$-flat, we have, for every $x\in \Support{\Dist}$,\footnote{Indeed, $1 = \sum_{x\in \Support{\Dist}} \Dist[x] \leq \Size{\Support{\Dist}}\cdot \max_{x\in \Support{\Dist}}  \Dist[x] \leq \Size{\Support{\Dist}}\cdot (1+\alpha)\min_{x\in \Support{\Dist}}  \Dist[x]$, and similarly
    $1 \geq \Size{\Support{\Dist}}\cdot \frac{\max_{x\in \Support{\Dist}}}{1+\alpha}  \Dist[x]$, from which 
    \[
    \frac{1}{(1+\alpha)\Size{\Support{\Dist}}} \leq \min_{x\in \Support{\Dist}} \Dist[x]\leq \max_{x\in \Support{\Dist}} \Dist[x] \leq \frac{1+\alpha}{\Size{\Support{\Dist}}}\,.
    \]}
    \[
        \frac{1}{(1+\alpha)\Size{\Support{\Dist}}} \leq \Dist[x] \leq \frac{1+\alpha}{\Size{\Support{\Dist}}}
    \]
    The probability that $x\samples \Dist$ belongs to $\Claimed{\Support{\Dist}}$ is then
    \begin{align*}
    \PProb{x\in \Claimed{\Support{\Dist}}}{x\samples \Dist}
    &= \Dist(\Claimed{\Support{\Dist}})
    = \sum_{y\in \Claimed{\Support{\Dist}}\cap \Support{\Dist}} \Dist[y] \\
    &\leq \Size{\Claimed{\Support{\Dist}}\cap \Support{\Dist}}\cdot \frac{1+\alpha}{\Size{\Support{\Dist}}} \\
    &\leq (1+\alpha)\frac{\Size{\Claimed{\Support{\Dist}}}}{\Size{\Support{\Dist}}} \\
    &\leq (1+\alpha)\frac{B}{B'}\,,
    \end{align*}
    and so the probability that all $s_2$ draws $x_1\dots,x_{s_2} \iidSamples \Dist$ fall in $\Claimed{\Support{\Dist}}$ is at most
  \begin{align*}
    \Prob{\TesterFunc \text{ outputs } \Accept} &= \PProb{\forall i \in [s_2], \,\, x_i \in \Claimed{\Support{\Dist}}}{x_1, \ldots, x_{s_2}\samples \Dist}\\ 
    &\leq \left((1+\alpha)\frac{B}{B'}\right)^{s_2} \le \delta_s\,, 
  \end{align*}
  the last inequality from our choice of $s_2 \geq \frac{\log(1/\delta_s)}{\log((1+\alpha)B'/B)}$.
    \bgroup
    \egroup
     \end{itemize}
     This concludes the proof.
\end{proof}

\subsection{Learning The Neighbourhood Of A ``Relevant'' Domain Element}\label{sec:onebucklearner}

\begin{definition}\label{defn:neighborhood}
Given a parameter $\kappa\in(0,1]$ and point $y\in \Domain$, let the \emph{$\kappa$-neighbourhood of $x$ under $\Dist$} be defined as
\[
	U^{\Dist}_\kappa(x) = \left\{ y \in \Domain : \frac{1}{1+\kappa }\TrueDist{x} \leq \TrueDist{y} \leq (1+\kappa)\TrueDist{x}\right\}
\]
that is, the set of points whose probability is within an $1+\kappa$ factor of the probability of $x$.	
\end{definition}

Later in the main protocol, the prover will claim that certain bucket indices of the $(\DomainSize, \Proximity)$-histogram of $\Dist$ have large mass. 
If the prover is honest, then buckets with large mass must also contain domain elements with heavy neighbourhoods (see claim \ref{claim:exists_y_star}), where the definition of neighbourhood is given in Definition \ref{defn:neighborhood}.
In an ideal world, we would like to verify this exact claim that the a domain element has a heavy $\kappa$-neighbourhood. 
However, due to boundary conditions and the inherent randomness of sampling, we are only able to verify a fuzzy version of the above claim.
More specifically, using a result by \citet[Lemma 3]{canonne2015testing} we show that, if the prover claims that the $\kappa$-neighbourhood (shown in \textcolor{blue}{blue} in Figure \ref{subfig:approximate-nbrhood}) of $x$ is heavy, we estimate the $\alpha_t \in [\kappa, 2\kappa]$ neighbourhood (shown in \textcolor{green!70!black}{green} in Figure \ref{subfig:approximate-nbrhood}) of $x$, up to a multiplicative factor of $\eta$.
This fuzziness is unavoidable, but we later show that this fuzzy approximation suffices to catch a cheating prover.
A distinctive feature of the $\alpha_t$-neighbourhood above is that apart from being heavy, it also has what we refer to as a ``moat'' (as shown in Figure \ref{subfig:moat})- by which we mean that most of the neighbourhood mass is concentrated away from the boundary.
This moat proves to be useful, when using the \nameref{lemma:pcond-empirical-guarantee} to check whether a sample $z\samples \Dist$ is in $\NeighbourhoodExpanded{x}{\alpha_t}$ or not. 
The idea is that the \texttt{Compare} algorithm inherently has some randomness we cannot avoid, and thus a sample that is actually in the moat but not in $\NeighbourhoodExpanded{x}{\alpha_t}$ might get incorrectly misclassified as being in the $\alpha_t$-neighbourhood. 
When the probability mass of the moat is small, the chances of sampling an element in the moat are slim, and therefore we are unlikely to err.
 See \nameref{lemma:sampling-lemma} Lemma for details.

\begin{figure}[htbp]
\centering

\begin{subfigure}{0.48\textwidth}
\centering
\begin{tikzpicture}[scale=0.65]
\draw[->] (0,0) -- (10,0) node[right, scale=0.6] {$[0,1]$};

\def\xpos{5}
\fill[red] (\xpos,0) circle (3pt);
\node[below, scale=0.6] at (\xpos, -0.4) {$\TrueDist{x}$};

\def\leftk{3.5}
\def\rightk{6.5}
\def\leftat{2.8}
\def\rightat{7.2}
\def\lefttwok{2.2}
\def\righttwok{7.8}

\draw[very thick, orange!80!black] (\lefttwok, 2.0) -- (\righttwok, 2.0);
\draw[very thick, orange!80!black] (\lefttwok, 0) -- (\lefttwok, 2.0);
\draw[very thick, orange!80!black] (\righttwok, 0) -- (\righttwok, 2.0);
\node[above, orange!80!black, scale=0.6] at (5, 2.1) {$U^\Dist_{2\kappa}(x)$};

\draw[very thick, green!70!black] (\leftat, 1.3) -- (\rightat, 1.3);
\draw[very thick, green!70!black] (\leftat, 0.0) -- (\leftat, 1.3);
\draw[very thick, green!70!black] (\rightat, 0.0) -- (\rightat, 1.3);
\node[above, green!70!black, scale=0.6] at (5, 1.4) {$U^\Dist_{\alpha_t}(x)$};

\draw[very thick, blue] (\leftk, 0.6) -- (\rightk, 0.6);
\draw[very thick, blue] (\leftk, 0) -- (\leftk, 0.6);
\draw[very thick, blue] (\rightk, 0) -- (\rightk, 0.6);
\node[above, blue, scale=0.6] at (5, 0.6) {$U^\Dist_\kappa(x)$};

\draw (\leftk,-0.05) -- (\leftk,0.05);
\draw (\rightk,-0.05) -- (\rightk,0.05);
\draw (\leftat,-0.05) -- (\leftat,0.05);
\draw (\rightat,-0.05) -- (\rightat,0.05);
\draw (\lefttwok,-0.05) -- (\lefttwok,0.05);
\draw (\righttwok,-0.05) -- (\righttwok,0.05);

%
%
%
\fill[blue] (3.8,0) circle (2.5pt);
\fill[blue] (4.4,0) circle (2.5pt);
\fill[blue] (5.6,0) circle (2.5pt);
\fill[blue] (6.2,0) circle (2.5pt);

\fill[green!70!black] (3.0,0) circle (2.5pt);
\fill[green!70!black] (7.0,0) circle (2.5pt);

\fill[orange!80!black] (2.4,0) circle (2.5pt);
\fill[orange!80!black] (7.6,0) circle (2.5pt);

\fill[gray] (1.0,0) circle (2.5pt);
\fill[gray] (9.0,0) circle (2.5pt);


\end{tikzpicture}
\caption{A fuzzy approximation of the the provers claim.}
\label{subfig:approximate-nbrhood}
\end{subfigure}
\hfill
\begin{subfigure}{0.48\textwidth}
\centering
\begin{tikzpicture}[scale=0.7]
\draw[->] (0,0) -- (10,0) node[right, scale=0.6] {$[0,1]$};
\def\xpos{5}
\fill[red] (\xpos,0) circle (3pt);

\node[below, scale=0.6] at (\xpos, -0.1) {$\TrueDist{x}$};
\def\leftat{2.8}
\def\rightat{7.2}
\def\leftatplus{2.5}
\def\rightatplus{7.5}
\fill[green!30, opacity=0.7] (\leftat, 0) rectangle (\rightat, 1.8);
\draw[thick, green!70!black] (\leftat, 0) -- (\leftat, 1.8);
\draw[thick, green!70!black] (\rightat, 0) -- (\rightat, 1.8);
\node[green!70!black, scale=0.65] at (5, 0.9) { $U_{\alpha_t}^{(\Dist)} \geq \beta$};
\fill[red!40, opacity=0.7] (\leftatplus, 0) rectangle (\leftat, 1.8);
\fill[red!40, opacity=0.7] (\rightat, 0) rectangle (\rightatplus, 1.8);
\draw[thick, red!70] (\leftatplus, 0) -- (\leftatplus, 1.8);
\draw[thick, red!70] (\rightatplus, 0) -- (\rightatplus, 1.8);
\draw[->, red!70, thick] (2.65, 2.35) -- (2.65, 1.9);
\draw[->, red!70, thick] (7.35, 2.35) -- (7.35, 1.9);
\draw[thick, purple!70!black] (\leftatplus, 0) -- (\leftatplus, 1.8);
\draw[thick, purple!70!black] (\rightatplus, 0) -- (\rightatplus, 1.8);
\node[left, scale=0.75] at (\leftatplus, 0.35) {$\frac{\TrueDist{x}}{1+\alpha_{t+1}}$};
\node[right, scale=0.75] at (\leftat, 0.35) {$\frac{\TrueDist{x}}{1+\alpha_t}$};
\node[left, scale=0.65] at (\rightat, 0.30) {$(1+\alpha_t)\TrueDist{x}$};
\node[right, scale=0.65] at (\rightatplus, 0.30) {$(1+\alpha_{t+1})\TrueDist{x}$};
\node[scale=0.7] at (5, 2.7) {$\TrueDist{\NeighbourhoodExpanded{x}{\alpha_{t+1}} \setminus \NeighbourhoodExpanded{x}{\alpha_{t}}} \leq \frac{\eta\beta}{32}$};
\end{tikzpicture}
\caption{A neighbourhood with a moat}
\label{subfig:moat}
\end{subfigure}
\caption{The prover claims that the mass of the $\kappa$-neighbourhood of $x$ is at least as large as $\beta$. 
The Estimate Neighbourhood algorithm of \citet{canonne2015testing} allows us to estimate $\alpha_t$ neighbourhood of $x$ up to a multiplicative factor of $\eta$, where $\alpha_t \in [\kappa, 2\kappa]$. 
Furthermore, $\alpha_t$ is such that while $\TrueDist{\NeighbourhoodExpanded{x}{\alpha_t}} > \beta$, there is a ``moat'' around the $\NeighbourhoodExpanded{x}{\alpha_t}$ such that $\TrueDist{\NeighbourhoodExpanded{x}{\alpha_{t+1}} \setminus \NeighbourhoodExpanded{x}{\alpha_{t}}} \leq \frac{\eta\beta}{32}$. 
For any sample $z \samples \Dist$, if we wish to use the \nameref{lemma:pcond-empirical-guarantee} to check whether $z \in \NeighbourhoodExpanded{x}{\alpha_t}$, this ``moat'' will act as insurance against the randomness of the $\PCond$ oracle, and therefore ensure our check is accurate with high probability.
}
\label{fig:moat}
\end{figure}

\begin{lemma}[{The \texttt{EstimateNeighborhood} procedure of \cite[Lemma~3]{canonne2015testing}}]\label{lemma:estimate-neighborhood}
	Given $\PCond$ access to $\Dist \in \DistSet{\Domain}$, there exists an algorithm \texttt{EstimateNeighborhood} which, on input $\kappa,\beta,\eta,\delta \in (0,1/2]$) as well as a domain element $x \in \Domain$, has the following guarantees. 
    Letting $T = 128/(\eta\beta\delta)$, and defining, for $t\in [T+1]$,
		\[
			\alpha_t = \kappa+(t-1)\kappa/T \in [\kappa,2\kappa)\,,
		\]
		the algorithm outputs a pair $(\hat{w}, t)\in [0,1]\times [T]$, such that $t$ is uniformly distributed in $[T]$, and:
		\begin{enumerate}
			\item Completeness: if $\Dist[U^{\Dist}_{\alpha_t}(x)] \geq \beta$, then
            \begin{align}
                \frac{\hat{w}}{\Dist[U^{\Dist}_{\alpha_t}(x)]} \in \left[\frac{1}{1+\eta}, 1+\eta\right] \quad&\text{and}\quad \TrueDist{\NeighbourhoodExpanded{x}{\alpha_{t+1}}\setminus \NeighbourhoodExpanded{x}{\alpha_t}} \leq \frac{\eta\beta}{32}
            \end{align}
			      
			      simultaneously with probability at least $1-\delta$.

			\item Soundness: if $\Dist[U^{\Dist}_{\alpha_t}(x)] < \beta$,
			      then
                  \begin{align}
                \hat{w} \leq (1+\eta)\beta \quad&\text{and}\quad \TrueDist{\NeighbourhoodExpanded{x}{\alpha_{t+1}}\setminus \NeighbourhoodExpanded{x}{\alpha_t}} \leq \frac{\eta\beta}{32}
            \end{align}
			       
			      simultaneously with probability at least $1-\delta$.
		\end{enumerate}
		The query complexity of the algorithm is $\Tilde{O}\left(\frac{1}{\kappa^2\eta^4\beta^3\delta^2}\right)$.
\end{lemma}
\begin{proof}
	This is a consequence of the guarantees from~\cite[Lemma~3]{canonne2015testing} (slightly simplified, and replacing $\eta$ by $\eta/2$ to have $[1/(1+\eta),1+\eta]$ instead of $[1-\eta,1+\eta]$ in the completeness guarantee).
\end{proof}

\begin{remark}
Note that, in the above statement, we could weaken the conditions of completeness and soundness to $\Dist[U^{\Dist}_{\kappa}(x)] \geq \beta$ and $\Dist[U^{\Dist}_{2\kappa}(x)] < \beta$, respectively, as $\Dist[U^{\Dist}_{\kappa}(x)] \leq \Dist[U^{\Dist}_{\alpha_t}(x)] \leq \Dist[U^{\Dist}_{2\kappa}(x)]$ for all $t\in[T]$.    
\end{remark}

\paragraph{Sampling from the neighbourhood.} Foreshadowing ahead, as a sub-routine of our main protocol, we will invoke the support size protocol on the re-normalised distribution $\Dist'$, which is obtained by restricting the domain of $\Dist$ to the set $\NeighbourhoodExpanded{x}{\alpha_t}$, and re-normalising $\TrueDist{\NeighbourhoodExpanded{x}{\alpha_t}}$.
In order to run the support size protocol as prescribed, it is important that the verifier be able to sample from $\Dist'$.
We now explain how after running \texttt{EstimateNeighbourhood} to obtain a tuple $(\hat{w},t)$ corresponding neighbourhood $U_t \Def \NeighbourhoodExpanded{x}{\alpha_t}$ and an estimate $\hat{w}$ of its probability mass, we can obtain \emph{samples} from $\Dist'$.


\begin{lemma}[Sampling From Neighbourhoods With Moats]\label{lemma:sampling-lemma}
Given parameters $\kappa, \beta, \delta, \eta \in (0,1/2)$ and domain element $x \in \Domain$. Setting $T = 128/(\eta\beta\delta)$ and let
\[
(\hat{w},t) \samples \texttt{EstimateNeighbourhood}(x;\kappa,\beta,\eta,\delta) 
\quad\text{and}\quad
\alpha_t = \kappa+(t-1)\kappa/T \in [\kappa,2\kappa).
\]
If $\Dist[U^{\Dist}_{\alpha_t}(\True{x})] \geq \beta$ and $\Dist[U^{\Dist}_{\alpha_{t+1}}(\True{x})\setminus U^{\Dist}_{\alpha_t}(\True{x})] \leq \frac{\eta\beta}{32}$\footnote{These are the guarantees obtained with high probability when $(\hat{w},t)$ is returned by \texttt{EstimateNeighbourhood} invoked on $\True{x}$ with parameters $\kappa,\beta,\eta,\delta$.}, then there exists a sampling algorithm with sample complexity $\BigO{\frac{\log(1/\eta)}{\beta}}$ and query complexity $\BigOTilde{\frac{1}{\beta^3\eta^2\delta^2\kappa^2}}$ that returns:
\begin{itemize}
\item $y \samples \Dist'$ with probability at least $1-\eta\log(1/\eta)$, where $\Dist'[y] \Def \frac{\TrueDist{y}}{\sum_{y' \in \NeighbourhoodExpanded{x}{\alpha_t}}\TrueDist{y'}}$, or
\item $\bot$ otherwise.
\end{itemize}
\end{lemma}

\begin{proof}
Suppose that $(\hat{w},t)\in [0,1]\times [T]$ is such that $\Dist[U^{\Dist}_{\alpha_t}(\True{x})] \geq \beta$ and $
                \Dist[U^{\Dist}_{\alpha_{t+1}}(\True{x})\setminus U^{\Dist}_{\alpha_t}(\True{x})] \leq \frac{\eta\beta}{32}
                $
both hold. 
That is, the neighbourhood is large enough (mass at least $\beta$), and it is surrounded by a ``moat'' (of probability mass at most $O(\eta\beta)$). 
The algorithm proceeds as follows, where we set $\delta' = c\cdot \eta\log(1/\eta)$ for a suitably small constant $c>0$.
\begin{itemize}
    \item Sample $s = \left\lceil\frac{\log(3/\delta')}{\log\frac{1}{1-\beta}}\right\rceil = O(\log(1/\delta')/\beta)$ points $x_1,\dots,x_s \iidSamples\Dist$. Note that (1) with probability at least $1-\frac{\delta'}{3}$ at least one $x_i$ will be in $U_t \Def U^{\Dist}_{\alpha_t}(\True{x})$, and (2) (by a union bound) with probability at least $1-\frac{s\eta\beta}{32} = 1-O(\eta\log(1/\delta'))$, no $x_i$ will be in the ``moat'' $U_t^{\parallel} \Def U^{\Dist}_{\alpha_{t+1}}(\True{x})\setminus U^{\Dist}_{\alpha_t}(\True{x})$.
    \item For each $x_i$, \texttt{Compare} on $\True{x},x_i$ with failure probability $\frac{\delta'}{3s}$, maximum ratio bound $K=2$, and accuracy parameter $\gamma = \BigTheta{\frac{\kappa}{T}}$
    to obtain an estimate of $\frac{\Dist[x_i]}{\Dist[x]}$ accurate to within an additive $\gamma$ (or detect that the ratio is not even in $[1/2,2]$). By a union bound over all $s$ calls to \texttt{Compare}, with probability at least $1-\frac{\delta'}{3}$ all estimates are indeed this accurate. By our choice of $\gamma$, this means that we can then decide which of the $x_i$'s are in $U_t$, and only (possibly) make a mistake on those which are in the moat $U_t^{\parallel}$. (This has query complexity $s\cdot \BigO{\frac{\log(s/\delta')}{\gamma^2}} = \BigO{\frac{\log(1/\delta')\cdot \log(s/\delta')}{\beta\cdot (\kappa/T)^2}}= \BigOTilde{\frac{1}{\beta^3\eta^2\delta^2\kappa^2}}$.)
    \item Return any element $x_i$ detected to be in $U_t$ uniformly at random, and denote it with $y$. Return $\bot$ if there is none.
\end{itemize}
By a union bound, except with probability at most 
\[
 \frac{\delta'}{3}+\frac{\delta'}{3}+\frac{s\eta\beta}{32}
 = \frac{2}{3}\delta'+ \BigO{\eta\log\frac{1}{\delta'}}
 \leq \delta'
\]
(the last inequality for $\frac{s\eta\beta}{32} \leq \frac{\delta'}{3}$, which is satisfied as $\delta' = \Omega(\eta\log(1/\eta))$), then 
we have simultaneously (1) at least one $x_i$ in $U_t$, (2) none in the moat $U_t^{\parallel}$, and (3) a correct decision on all elements (since the only errors would be for elements in $U^{\parallel}$, and there aren't any by (2)). That is, with probability at least $1-\delta'$, the above procedure returns an element $y$ sampled from $\Dist'$.
\end{proof}

\section{Interactive Proofs For Label Invariant Properties}\label{sec:label-invariance}

Given parameter $\HistFactor$, in this section, we describe our main result, an interactive proof system for certifying the $\Proximity$-approximate histogram, where $\Proximity \Def 0.01\HistFactor$.  
Note that the histogram is constructed at a finer granularity than the proximity $\HistFactor$ parameter warrants, as $\Proximity < \HistFactor$, but this only makes testing for label invariant properties with proximity parameter $\HistFactor$ easier.

\begin{mdframed}[
    frametitle={The Approximate Single Algorithm},
    frametitlealignment=\centering,
]

\noindent\textbf{Verifier Input:}
\begin{itemize}[nosep,leftmargin=1.5em]
    \item Domain $\Domain$, soundness and completeness parameters $\delta' \in (0,1)$, proximity parameter $\Proximity \in (0, 0.1)$ and $\PCond$ access to $\Dist$ 
    \item Prover claims: $\Claimed{\BucketOf{\Proximity}{\yStar}} = \jStar$ and 
    $\ClaimedDist{\NeighbourhoodExpanded{\yStar}{\Proximity/3}} \ge \frac{\Proximity}{30\NumBuckets}$
\end{itemize}

\vspace{6pt}
\noindent\textbf{Verifier Output:} Either $\Reject$ or $\Accept$ of above claim.

\vspace{8pt}
\noindent\hrulefill
\vspace{8pt}

\noindent\textbf{Parameters For Invoking Sub-routines}
\vspace{4pt}
\begin{itemize}[nosep,leftmargin=1.5em]
    \item Set $\delta_c = \delta_s = \delta'/2$. \\
    \item Set $\kappa = \Proximity/3$, $\beta = \frac{\Proximity}{30\NumBuckets}$. \\
    \item Set $\eta = \min\left(\frac{\Proximity}{10}, \frac{\log s_1}{100s_1\delta'}, \frac{\log s_2}{100s_2\delta'}\right) = \BigOmegaTilde{\Proximity^4/\delta'}$ where $s_1,s_2$ are sample complexities from Figures \ref{alg:small-supp-size} and \ref{alg:large-supp-size} with $B'/B = A/A' = 1+\Proximity$.
    
\end{itemize}
 
\vspace{8pt}
\noindent\hrulefill
\vspace{8pt}

\noindent\textbf{\large Algorithm Body}

\vspace{8pt}
\noindent\textbf{Step 1: Estimate Neighborhood}

\vspace{4pt}
Run $(\hat{w}, t) \gets \texttt{EstimateNeighbourhood}(\yStar; \kappa, \eta, \beta, \delta'/2)$

\textbf{If} $\hat{w} \le (1 + \eta)\beta$ \textbf{then return} $\Reject$. 

Otherwise, set $T = 128/(\eta\beta\delta)$ and $\alpha \Def \kappa+(t-1)\kappa/T \in [\kappa,2\kappa)$.

\vspace{10pt}
\noindent\textbf{Step 2: Compute Bounds}

\vspace{4pt}
Set:
\begin{align*}
    A' &= \frac{\hat{w} \DomainSize}{\Proximity(1 + \Proximity)^{\jStar+1}(1 + \eta)(1+\alpha)}, \quad
    A = \frac{\hat{w}\DomainSize}{\Proximity(1 + \Proximity)^{\jStar}(1 + \eta)(1+\alpha)} \\[8pt]
    B &= \frac{\hat{w}(1 + \eta)(1+\alpha)\DomainSize}{\Proximity(1 + \Proximity)^{\jStar-1}}, \quad
    B' = \frac{\hat{w}(1 + \eta)(1+\alpha)\DomainSize}{\Proximity(1 + \Proximity)^{\jStar-2}}
\end{align*}

Define $\Dist' \in \DistSet{\NeighbourhoodAlpha{\yStar}}$ where $\Dist'[x] = \frac{\TrueDist{x}}{Z}$ with $Z = \sum_{x' \in \NeighbourhoodAlpha{\yStar}} \TrueDist{x'}$.

\vspace{10pt}
\noindent\textbf{Step 3: Support Size Interactive Protocols}

\vspace{4pt}
\textbf{If} $B \in \BigO{\sqrt{\DomainSize}}$ \textbf{then}

\quad Run Figure \ref{alg:small-supp-size} with parameters $A', A, B, B'$, $\delta_c = \delta_s = \delta'/2$.

\textbf{Else if} $A \in \SmallOmega{\sqrt{\DomainSize}}$ \textbf{then}

\quad Run Figure \ref{alg:large-supp-size} with parameters $A', A, B, B'$, $\delta_c = \delta_s = \delta'/2$.\\[4pt]

In both cases sample from $(1+\alpha)$-flat distribution $\Dist'$ using Lemma \ref{lemma:sampling-lemma} with $\kappa = \Proximity/3$.

\textbf{If} either algorithm outputs $\Reject$ \textbf{then return} $\Reject$.

\vspace{10pt}
\noindent\textbf{Return} $\ClaimedDist{\yStar} = \frac{\Proximity(1+\Proximity)^\jStar}{N}$ and $\Accept$.

\vspace{6pt}
\captionof{figure}{Approximating the probability mass of a single domain item.}
\label{alg:approximate-single}

\end{mdframed}

Recall, for any distribution $\Dist \in \DistSet{\Domain}$ and domain element $z \in \Domain$, we use $\BucketOf{\Proximity}{z} \in \BucketIndices$ to denote the bucket index of the $(\DomainSize, \Proximity)$-histogram of $\Dist$ that $z$ belongs to. 

\begin{theorem}[Approximate Single]\label{lemma:approximate-single}
  Fix $\Proximity > 0$ and $\Dist \in \DistSet{\Domain}$ for some large enough $\DomainSize \in \Naturals$.
	Let $\NumBuckets$ denote the number of buckets of the $(\DomainSize, \Proximity)$-histogram of $\Dist$, 
	The protocol starts with the provers claim that for domain element $\yStar \in \Domain$, its bucket index is $\Claimed{\BucketOf{\Proximity}{\yStar}}=\jStar$, and additionally that $\yStar$ has a heavy $\Proximity/3$-neighbourhood i.e. $\ClaimedDist{\NeighbourhoodExpanded{\yStar}{\Proximity/3}} \ge \frac{\Proximity}{30\NumBuckets}$.
	For every  $\delta' \in (0,1/2)$, the proof system described by Figure \ref{alg:approximate-single}, causes the verifier to output $\Accept$ or $\Reject$ based on the following properties:
	\begin{enumerate}

		\item \textbf{Completeness:} If the prover is honest, then 
        \begin{align*}
\left(\Tag{\yStar}=\True{\BucketOf{\Proximity}{\yStar}}\right) &\land             \left(\TrueDist{\NeighbourhoodExpanded{\yStar}{\Proximity/3}} = \ClaimedDist{\NeighbourhoodExpanded{\yStar}{\Proximity/3}} \ge \frac{\Proximity}{30\NumBuckets}\right)
\end{align*}
As a result $\Prob{ \TesterFunc \text{ outputs } \Reject}  \le \delta'$

\item{ \textbf{Soundness:} For any cheating prover strategy if 
\begin{align*}
\left(\,\left|\Tag{\yStar}-\True{\BucketOf{\Proximity}{\yStar}}\right| \ge 2\,\right) &\lor \left( \TrueDist{\NeighbourhoodExpanded{\yStar}{2\Proximity/3}} \le \frac{\Proximity}{30\NumBuckets}\right)
\end{align*}
then$\Prob{ \TesterFunc \text{ outputs } \Accept } \le \delta'$
		}
	\end{enumerate}

If the verifier outputs $\Accept$, then it accepts that $\ClaimedDist{\yStar} = \frac{\Proximity(1+\Proximity)^{\jStar}}{\DomainSize}$ as the approximate probability mass of $\yStar$ under $\Dist$.
\end{theorem}

\begin{proof}
Set $T = 128/(\eta\beta\delta)$.
Re-stating the parameters defined in Figure \ref{alg:approximate-single}, we have $\delta_c = \delta_s = \delta'/2$, $\kappa = \Proximity/3$, $\beta = \frac{\Proximity}{30\NumBuckets}$, and $\eta = \min\left(\frac{\Proximity}{10}, \frac{\log s_1}{100s_1\delta'}, \frac{\log s_2}{100s_2\delta'}\right) = \BigOmegaTilde{\Proximity^4/\delta'}$ where $s_1,s_2$ are sample complexities from Figures \ref{alg:small-supp-size} and \ref{alg:large-supp-size} with $B'/B = A/A' = 1+\Proximity$. 
Let 
\[(\hat{w}, t)\samples\texttt{EstimateNeighborhood}(\yStar; \Proximity/3, \eta, \beta, \delta'/2)\]
and 
$\alpha \Def \kappa +(t-1)\kappa/T\in [\Proximity/3, 2\Proximity/3]$.\\

\textbf{Completeness}: As the prover is honest, we have $\jStar = \BucketOf{\Proximity}{\yStar}$, and therefore,

	\begin{equation}
		\frac{\ClaimedDist{\yStar}}{\TrueDist{\yStar}} \in [\frac{1}{1+\Proximity} , 1+\Proximity ]
	\end{equation}

where $\ClaimedDist{\yStar} = \frac{\Proximity(1+\Proximity)^{\jStar}}{\DomainSize}$.
 Additionally, we also have $ \TrueDist{\NeighbourhoodExpanded{\yStar}{\alpha}} \ge \TrueDist{\NeighbourhoodExpanded{\yStar}{\Proximity/3}} \ge \beta \Def \frac{\Proximity}{30\NumBuckets}$.\par
 
As we are upper bounding the completeness error, we assume that any time $\hat{w}$ is \emph{not} a good approximation of $\TrueDist{\NeighbourhoodAlpha{\yStar}}$, the worst happens, the verifier rejects the honest provers claim.
$\hat{w}$ is a bad approximation with probability at most $\delta'/2$.
Finally, assuming that the verifier does not reject the honest provers claim, then this restricts the size of $\NeighbourhoodAlpha{\yStar}$
More specifically, from the provers claim $\jStar$ is the bucket of $\yStar$, we get an upper and lower bound on $\Size{\NeighbourhoodAlpha{\yStar}}$.
More specifically 

\begin{align}
    \Size{\NeighbourhoodAlpha{\yStar}} &\le \frac{\TrueDist{\NeighbourhoodAlpha{\yStar}}}{\min_{z \in \NeighbourhoodAlpha{\yStar}}\TrueDist{z}} \\
    &\le \frac{\TrueDist{\NeighbourhoodAlpha{\yStar}}(1+\alpha)}{\TrueDist{\yStar}} \label{eq:alpha-nbrhodd}\\
    &\le \frac{\TrueDist{\NeighbourhoodAlpha{\yStar}}(1+\alpha)\DomainSize}{\Proximity(1 + \Proximity)^{\jStar -1}} \label{eq:bucket-limits}\\
    &\le \frac{\hat{w}(1+\eta)(1+\alpha)\DomainSize}{\Proximity(1 + \Proximity)^{\jStar -1}} \label{eq:approx-estimate-nbrhood}
\end{align}

where Equation \eqref{eq:alpha-nbrhodd} comes from the definition of a $\alpha$-neighbourhood,\eqref{eq:bucket-limits} comes from the limits of bucket index $\jStar$, and \eqref{eq:approx-estimate-nbrhood} comes from the approximation guarantees of the \nameref{lemma:estimate-neighborhood}. A lower bound on the size of $\NeighbourhoodAlpha{\yStar}$ can similarly defined, giving us
\begin{align}
	\Size{\NeighbourhoodAlpha{\yStar}} &\le \frac{(1 + \alpha) \DomainSize\,\,\TrueDist{\NeighbourhoodAlpha{\yStar}}}{\Proximity(1 + \Proximity)^{\jStar-1}} \le \frac{ \DomainSize\,\,\hat{w}(1+\eta)(1 + \alpha)}{\Proximity(1 + \Proximity)^{\jStar-1}} \Def B \\
	\Size{\NeighbourhoodAlpha{\yStar}} &\ge \frac{ \DomainSize\,\,\TrueDist{\NeighbourhoodAlpha{\yStar}}}{(1 + \alpha)\Proximity(1 + \Proximity)^{\jStar}} \ge \frac{\hat{w} \DomainSize}{\Proximity(1 + \Proximity)^{\jStar}(1 + \eta)(1+\alpha)} \Def A 	
\end{align}

Define $\Dist' \in \DistSet{\NeighbourhoodAlpha{\yStar}}$ such that for all $x \in \NeighbourhoodAlpha{\yStar}$, $\Dist'[x] = \frac{\TrueDist{x}}{Z}$
, where $Z \Def \sum_{x' \in \NeighbourhoodAlpha{\yStar}} \TrueDist{x'}$.
That is $\Dist'$ is $\Dist$ restricted and re-normalised on $\NeighbourhoodAlpha{\yStar}$.
As the verifier has not yet rejected, from the guarantees of the Lemma~\ref{lemma:estimate-neighborhood} we are in the regime where $\TrueDist{\NeighbourhoodAlpha{\yStar}} \ge \beta$ and the mass of the ``moat'' is at most $\frac{\eta\beta}{32}$.
This implies that we can invoke the \nameref{lemma:sampling-lemma} to sample from $\Dist'$.
More specifically, to obtain one sample $z \samples \Dist'$ we need to draw $\BigOTilde{\frac{\log 1/\eta}{\beta}}$ samples from $\Dist$, and make $\BigOTilde{\frac{1}{\beta^3 \eta^2\delta'^2\kappa^2}}$ $\PCond$ queries, and with probability $1- \eta \log (1/\eta)$ we succeed in obtaining this sample.
We want to to invoke the support size protocol with $A, A'$ and $B, B'$ as defined in Figure \ref{alg:approximate-single}.
In total we need to draw $\BigO{\left(1/\Proximity\right)^4 \log \frac{1}{\delta'}}$ samples from $\Dist'$.

The probability that the support size protocol rejects is at most $\delta'/2$.
Thus, the total completeness error for this protocol is $\delta'$.

\textbf{Soundness}: Soundness follows pretty similarly.
Assume that the prover lies about by more than 1 bucket, then 
	\begin{equation}
		\frac{\ClaimedDist{\yStar}}{\TrueDist{\yStar}} \notin [\frac{1}{(1+\Proximity)^2} , (1+\Proximity)^2]
	\end{equation}

Let $\beta$ be as defined in Figure \ref{alg:approximate-single}.
If the prover \emph{also} lied about the mass of the neighbourhood around $\yStar$, and 
$\TrueDist{\NeighbourhoodExpanded{\yStar}{2\Proximity/3}} \le \beta$, then the probability that the prover does not output $\Reject$ is at most $\delta'/2$.
Assuming, $\TrueDist{\NeighbourhoodExpanded{\yStar}{2\Proximity/3}} \ge \beta$, the the probability that $\hat{w}$ is not a good approximation of $\TrueDist{\NeighbourhoodAlpha{\yStar}}$ is at most $\delta/'2$.
By the worst case assumption, the verifier accepts the provers claim with probability at most $\delta'/2$.
Assuming that $\hat{w}$ is a good approximation, and $\TrueDist{\NeighbourhoodExpanded{\yStar}{2\Proximity/3}} \ge \beta$, once again, by a similar analysis, this implies that $\Size{\NeighbourhoodAlpha{\yStar}} \ge B'$ or $\Size{\NeighbourhoodAlpha{\yStar}} \le A'$, where 
\begin{align}
B' &= \frac{\hat{w}(1 + \eta)(1+\alpha)\DomainSize}{\Proximity(1 + \Proximity)^{\jStar-2}}\\
A' &= \frac{\hat{w}\DomainSize}{\Proximity(1 + \Proximity)^{\jStar+1}(1 + \eta)(1+\alpha)}	
\end{align}

$\Dist'$ is $(1+\alpha)$-flat and sample-able, so we get from soundness of the support size protocol, the prover accepts with probability at most $\delta'/2$.

\textbf{Complexity}: For $\texttt{EstimateNeighbourhood}$ the verifier makes $\Tilde{O}\left(\frac{\NumBuckets^3}{\Proximity^5 \eta^4 \delta'^2}\right) = \Tilde{O}\left(\frac{(\log N)^3}{\Proximity^{24} \delta'^2}\right)$ queries and samples.
For $\texttt{SupportSize}$ we draw at worst $\BigO{\left(1/\Proximity\right)^4 \log \frac{1}{\delta'}}$ samples (from the distribution restricted to the neighborhood), and make at worst $\BigOTilde{\left(1/\Proximity\right)^4\log \frac{1}{\delta'}}$ $\PCond$ queries. Substituting in the sample complexity of generating a single sample from the neighborhood (Lemma~\ref{lemma:sampling-lemma}) This corresponds to $\BigOTilde{L_{\tau} \cdot \left(1/\Proximity\right)^5 \log \frac{1}{\delta'}} = \BigOTilde{\log N \cdot \left(1/\Proximity\right)^6 \log \frac{1}{\delta'}}$ samples from the true distribution, and $\BigOTilde{L_{\tau}^3 \cdot \left(1/\Proximity\right)^{21} \cdot \frac{1}{\delta'^2}} = \BigOTilde{(\log N)^3 \cdot \left(1/\Proximity\right)^{24} \cdot \frac{1}{\delta'^2}}$  $\PCond$ queries.
The communication complexity is at worst $\BigOTilde{\sqrt{\DomainSize}\poly(1/\Proximity)}$.
The round complexity is 
$\BigO{\left(1/\Proximity\right)^4 \cdot \log \frac{1}{\delta'}}$.

\end{proof}

\begin{remark}
In the case that the prover claims $\Tag{\yStar} \le 1$ or $\Tag{\yStar} \ge \NumBuckets-1$, then we are saved from doing 2 sided tests, and can just check one side for support size decision problem.
Of course if a bucket index large enough, and has large enough mass, then the verifier could simply find all the points in the bucket by mere sampling.
We do not explicitly spell out this easy case, as its covered by the protocol above.
\end{remark}

\subsection{Succinct Proof Systems For Any Label Invariant Property}

\begin{claim}\label{claim:exists_y_star}
Let $B$ be a bucket of the $(\DomainSize, \Proximity)$-histogram of a distribution $\Dist$ with probability mass $p$.
Then, there exists a point $\yStar \in B$ such that $\TrueDist{\NeighbourhoodExpanded{\yStar}{\Proximity/3}} \ge p/3$, where the neighborhood $\NeighbourhoodExpanded{\yStarJ}{\Proximity/3}$ is as defined in Definition \ref{defn:neighborhood}.
\end{claim}

\begin{proof}
For any $x \in B$, define $t(x) \Def -\log \TrueDist{x}$. 
From the definition of a $(\DomainSize, \Proximity)$-histogram we have that $\Delta \Def \max_{x \in B} t(x) - \min_{x\in B}t(x) \le \log(1+\Proximity)$.
Define $\delta = \log (1 + \Proximity/3)$.
 Let $S = x_1, \ldots, x_{\Size{B}}$ denote the ordered set of the points in $B$ in ascending order of their probability mass (ties broken arbitrarily).
 Now we greedily partition $S$ into disjoint clusters of mass at most $2\delta$.
 As the total mass of $S$ is $p$, the maximum number of clusters in $S$, denoted by $M$ is at most $\lceil \frac{\Delta}{2\delta}\rceil+1$.
By the averaging argument, there must be one cluster with mass at least $\frac{p}{M}$.
Let $\yStar$ denote the median point in this cluster. 
This implies that $\TrueDist{\NeighbourhoodExpanded{\yStar}{\alpha}} \ge \frac{p}{M}$ (as the total mass of the cluster is $2\delta$ and $\yStar$ is the median).
To get the desired lower bound, we upper bound $M \le \lceil \frac{\Delta}{2\delta}\rceil+1 \le 2$, where the last inequality comes from the fact that $\Delta = \log (1 + \Proximity) \le 3\log (1 + \Proximity/3) = 3\delta$, giving us $M \le 3$.

\end{proof}
	
\begin{mdframed}[
		frametitle={Proof System For Verifiable Succint Histograms},
		frametitlealignment=\centering
	]
	\textbf{Common Inputs}: $\HistFactor \in (0, 0.1)$. Verifier has $\PCond$ access to $\Dist$. Prover knows $\Dist$ in clear.\\

  Define $\Proximity \Def 0.01\HistFactor$, $\gamma = \frac{\Proximity}{s(1+\Proximity)}$, $s\Def \frac{10\NumBuckets}{\Proximity}\log (10\NumBuckets)$.\\

	\begin{enumerate}

		\item{\textbf{Verifier} $\leftarrow$ \textbf{Prover}:   For every ``relevant'' bucket, i.e bucket indices $j \in \BucketIndices$ such that $\TrueDist{B_j^{(\Dist, \Proximity)}}\ge \frac{\Proximity}{10\NumBuckets}$, send the verifier $y_j \in B_j^{(\Dist, \Proximity)}$ such that $\TrueDist{\NeighbourhoodExpanded{y_k}{2\Proximity/3}} \ge \frac{\Proximity}{30\NumBuckets}$, along with its bucket index $\Tag{y_j}$. That is the prover is asked to declare all the relevant bucket indices $j$, and a point $y_j$ in each relevant bucket with a heavy neighborhood. 
        Prover claims that $\Claimed{K} \subseteq \BucketIndices$ are the relevant buckets, and sends $y_1, \ldots, y_{\Size{\Claimed{K}}}$ as representative samples for each bucket.
		}

        \item{\textbf{Verifier} $\leftrightarrow$ \textbf{Prover}: For each $j \in \Claimed{K}$, Invoke the proof system specified by \nameref{lemma:approximate-single} with input $y_j$, parameters $\Proximity/3$, and $\delta = \frac{1}{20\Size{\Claimed{K}}}$, and output $\Reject$ if the proof system rejects \emph{any} of the tests. 
        Otherwise, accept the claim $\ClaimedDist{y_j} = \frac{\Proximity(1 + \Proximity)^j}{\DomainSize}$ for all $j \in \Claimed{K}$.}
	
		\item{\textbf{Verifier $\rightarrow$ Prover:} Send set $S = (x_1, \ldots, x_s) \iidSamples \Dist$.
		      }
		      
        \item{\textbf{Verifier} $\leftarrow$ \textbf{Prover} Send $\{\Tag{x_1}, \ldots, \Tag{x_s}\}$,
		      }

        	\item{\textbf{Verifier Computation:} Compute for all $j \in \BucketIndices$. $\Claimed{p_j} \Def \frac{1}{\Size{S}}\Size{\left\{x \in S: \Tag{x} =j \right\}}$
		      }
        
		 \item{
		 \textbf{Verifier Computation}: For each $x \in S$, let $\yStar(x) \in \{y_1, \ldots, y_{\Size{\Claimed{K}}}\}$ be the coreresponding $y_j$ such that $\Tag{\yStar(x)} = \Tag{x} = j$. 
		 If no just $\yStar(x)$ exists for even one $x$, the verifier outputs $\Reject$. The verifier also rejects if the prover declares multiple $\yStar$'s for a single $x$. Otherwise}\\
	\end{enumerate}		 
\begin{algorithmic}
\For{$x \in S$}
  \State Compute $\alpha_{x\yStar(x)} \gets \texttt{Compare}\!\left(\{x_i\}, \{\yStar(x_i)\}, \gamma, \tfrac{1}{20s}, K=2\right)$\\
\EndFor

\State If \texttt{Compare} outputs HIGH or LOW, for any $x$, output $\Reject$.\\

\State Define $\ClaimedDist{x} \Def  \frac{\Proximity(1+\Proximity)^{\Tag{x}}}{\DomainSize}$ for all $x \in S$ \\

  \If{%
    $\displaystyle 
      \sum_{x \in S} 
      \bigl| \ClaimedDist{\yStar}\,\alpha_{x\yStar(x)} - \ClaimedDist{x} \bigr|
      \;\;\ge\;\;
      2\HistFactor - (2\Proximity + \Proximity^2 )- \frac{\Proximity}{1+\Proximity}$%
  }
    \State \textbf{Output} $\Reject$ and halt
  \Else
    \State \textbf{Output} $\ClaimHistExpanded$ as the learned histogram
  \EndIf
\end{algorithmic}
    \captionsetup{hypcap=false} 
	\captionof{figure}{A General Proof System To Obtain Verifiable $(\DomainSize, \Proximity)$-Histograms for any distribution $\Dist \in \DistSet{\Domain}$} \label{alg:label-invariant-property}
    \captionsetup{hypcap=true} 
\end{mdframed}

\begin{theorem}[Public-Coin Proof System For Succinct Histograms]\label{thm:main-thm}
	Fix $\DomainSize \in \Naturals$, $\HistFactor \in (0, 0.1)$ and  $\Dist \in \DistSet{\Domain}$.
	Define $\Proximity \Def 0.1 \HistFactor$.
  There exists an interactive protocol $\Protocol$ between an honest verifier $\TesterFunc$, and an omniscient untrusted prover $\Prover{\Dist}$, described in Figure \ref{alg:label-invariant-property}, where the verifier has $\PCond$ access to $\Dist$, such that at the end of the interaction the verifier either outputs a $\ApproxHistParam$ histogram $\{\Claimed{p_j}\}_{j \in \BucketIndices}$ or outputs $\Reject$, with the following constraints:

	\begin{enumerate}
		\item{\textbf{Completeness:} If the prover follows the protocol as prescribed, then
		      \[ \Prob{\outputs{\ProofSystem{\Dist}{\HistFactor, \DomainSize}} = \Accept \land \RL{\Dist}{\ClaimHistExpanded} \le 0.1\HistFactor } \ge 2/3 \]

		      }
		\item{\textbf{Soundness:} For any cheating strategy $\ChProver{\Dist}$, we have

		      \[ \Prob{\outputs{\ChProofSystem{\Dist}{\HistFactor, \DomainSize}} \neq \Reject \land \RL{\Dist}{\ClaimHistExpanded} \ge \HistFactor } \le 1/3 \]

		      }
	\end{enumerate}

  \begin{enumerate}
    \item \textbf{Sample and Query Complexity}: $\Tilde{O}\left(\frac{(\log \DomainSize)^4}{\poly(\Proximity)}\right)$
    \item \textbf{Communication Complexity} $\BigOTilde{\sqrt{\DomainSize}\poly(1/\Proximity)}$
    \item \textbf{Round Complexity}: $\BigOTilde{\log \DomainSize \cdot \poly(1/\Proximity)}$
  \end{enumerate}
\end{theorem}

\begin{proof}

Let $K \subseteq \NumBuckets$ denote the indices of the ``relevant'' buckets of the $(\DomainSize, \Proximity)$-histogram of $\Dist$.
By Claim \ref{claim:exists_y_star}, for each relevant bucket $j \in K$ there exists some $y_j \in B_j^{(\Dist, \Proximity)}$ such that $\TrueDist{\NeighbourhoodExpanded{y_j}{\Proximity/3}} \ge \frac{\Proximity}{30L}$.

\textbf{Completeness}:
As the prover is honest, the histogram that the verifier outputs is simply the histogram one would learn if they could sample from the histogram itself. 
The folklore learning theorem says it is possible to learn an arbitrary distribution over domain of size $\NumBuckets$ up to $\epsilon$ error in total variation distance with probability $\delta$ takes $\BigO{\frac{\NumBuckets + \log (1/\delta)}{\epsilon^2}}$. 
Setting $\delta = 1/20$, $\epsilon = 0.1\HistFactor$, and from our definition of $s$, we have with probability at most 1/20 we fail to learn a good approximation of the histogram, with room to spare. 
Thus, with probability at least $1 - 1/20$ we have that $\RL{\Dist}{\ClaimHistExpanded} \le 0.1\HistFactor$.\\ 
Next we show that the probability that the prover outputs $\Reject$ is upper bounded by $1/3$.
Based on Claim \ref{claim:exists_y_star}, the honest prover is able to find a good $y_j$ for all $j \in K$ where $\Claimed{K} = K$.
We also have the guarantee that for all $j \in K$ (as the prover is honest about the bucket index of $y_j$)

	\begin{equation}
		r_j \Def \frac{\ClaimedDist{y_j}}{\TrueDist{y_j}} \in \left[\frac{1}{1 +\Proximity}, 1+\Proximity\right] \label{eq:good-approx}
    \end{equation}
    
For every $j \in K$, the verifier runs \texttt{Approximate Single} with $\yStar= y_j$ and $\delta' = \frac{1}{20L}$. 
Therefore, the probability that the verifier accidentally rejects any of the claims about $\ClaimedDist{y_j}$ is at most $1/20$ (the union bound over the probability that $\texttt{Approximate Single}$ fails).\\

Let $S$ be the set of samples as described in Figure \ref{alg:label-invariant-property}, with $s \Def \Size{S}$. 
For each $x \in S$, let $\Claimed{\BucketOf{\Proximity}{x}} \in \BucketIndices$ denote the provers claim about the bucket index of $x$.
If for any $x$, $\Claimed{\BucketOf{\Proximity}{x}} \notin \Claimed{K}$, the verifier outputs $\Reject$ immediately.
If the prover honestly declares all relevant buckets, this event happens when at least 1 out of the $s$ samples from $\Dist$ are in ``non-relevant'' buckets, which happens with probability at most $\Proximity/10 < 1/20$.

Set $\delta = 1/20$ and $\gamma = \frac{\Proximity}{s(1+\Proximity)}$. 
As we did not reject in the previous step, for each $x \in S$, there is a corresponding $y_k$ such that $k = \Claimed{\BucketOf{\Proximity}{x}}$.
This implies, with only $\BigO{\frac{K\log 2s/\delta}{\gamma^2}}$ $\PCond$ queries, the \nameref{lemma:pcond-empirical-guarantee} guarantees, for every $x \in S$, where $K=2$.

	\begin{equation}
		\PProb{\left|\alpha_{x y_k} - \frac{\TrueDist{x}}{\TrueDist{y_k}} \right| \ge \gamma }{\alpha_{x y_k} \samples \text{Compare}(x, y_k; \gamma, \nicefrac{\delta}{s}) } \le \frac{\delta}{s}
	\end{equation}

Let $\Claimed{\Dist} =\text{arg}\min_{D' \leftarrow \ClaimHistExpanded} \TV{\True{\Dist}}{D'}$,
	where we used the notation that $D' \leftarrow \ClaimHistExpanded$ to denote that the $(\DomainSize, \Proximity)$-histogram of $\Dist'$ is $\ClaimHistExpanded$.
In what follows, for any $x \in S$, we use $\yStar(x)$ to denote the corresponding $y_k$ such that $\Claimed{\BucketOf{\Proximity}{x}}=k$, $r(x)$ to be the corresponding ratio $r_k$.
Thus, we have with probability $1 - \delta/s$, for any given $x$,
\begin{align*}
    \left| \ClaimedDist{\yStar(x)}\alpha_{x\yStar} - \ClaimedDist{x} \right| & = \left| \ClaimedDist{\yStar(x)}\alpha_{x\yStar(x)} - \ClaimedDist{\yStar(x)}\frac{\TrueDist{x}}{\TrueDist{\yStar(x)}} +\ClaimedDist{\yStar(x)}\frac{\TrueDist{x}}{\TrueDist{\yStar(x)}} - \ClaimedDist{x} \right| \\
    &\le \ClaimedDist{\yStar(x)}\left| \alpha_{x\yStar(x)} - \frac{\TrueDist{x}}{\TrueDist{\yStar(x)}} \right| + \left| \ClaimedDist{\yStar(x)}\frac{\TrueDist{x}}{\TrueDist{\yStar(x)}} - \ClaimedDist{x} \right| \\
    &\le \ClaimedDist{\yStar(x)}\gamma + \left|\TrueDist{x}r(x) - \ClaimedDist{x} \right| \\
    &= \ClaimedDist{\yStar(x)}\gamma + \left|\TrueDist{x}(r(x)-1) + \TrueDist{x} - \ClaimedDist{x} \right| \\
    &\le \gamma + \TrueDist{x}\left|r(x)-1\right| + \left| \TrueDist{x} - \ClaimedDist{x}\right|\\
    &\le \gamma + \TrueDist{x}\cdot \tau + \left| \TrueDist{x} - \ClaimedDist{x}\right|
\end{align*}
where for the last inequality we used $|r(x)-1| \le \max\left\{1 - \frac{1}{1 + \Proximity}, 1 + \Proximity -1 \right\} \le \Proximity$.
Summing over all $x \in S$, and taking the union bound, with probability at least $1-1/20$ we have
\begin{align*}
 \sum_{x \in S} \left| \ClaimedDist{\yStar(x)}\alpha_{x\yStar(x)} - \ClaimedDist{x} \right| &\le s\gamma + \left|(r-1)\right|\sum_{x \in S}\TrueDist{x} + \sum_{x \in S}\left| \TrueDist{x} - \ClaimedDist{x}\right|\\
 & \le  3\Proximity \\
		 & < 2\HistFactor - (2\Proximity + \Proximity^2) -  \frac{\Proximity}{1 + \Proximity} \,.
\end{align*} 
Of course there is a chance that the prover does everything honestly, and the verifier accepts but the verifier got very unlucky, and the samples in $S$ are a bad empirical estimator for $\TrueHist$.
With how we set $s$, this happens with probability at most 1/20.
This completes the proof for completeness with error $4/20 < 1/3$.

\textbf{Soundness}: For any $\yStar \in \{y_1, \ldots, y_k\}$, if the prover lies about $\ClaimedDist{\yStar}$ such that $\frac{\ClaimedDist{\yStar}}{\TrueDist{\yStar}} \notin [\frac{1}{(1 + \Proximity)^2} , (1+\Proximity)^2 ]$, based on the soundness of Theorem \ref{lemma:approximate-single}, $\Prob{\TesterFunc \text{ outputs }\Accept} \le 1/20L $.
We require the verifier to not reject \emph{any} of the tests, so by a union bound, the probability that the verifier accepts given the prover grossly lied about any $y_j$'s bucket index, is at most $1/20$ with plenty of room to spare.

Conditioning on the event that that the verifier did not reject the claim about the mass of any $\yStar \in \{y_1, \ldots, y_k\}$, we have with probability at least $ 1 - 1/20$, for all $\yStar \in \{y_1, \ldots, y_k\}$,

	\[
		r \Def \frac{\ClaimedDist{\yStar}}{\TrueDist{\yStar}} \in [\frac{1}{(1 + \Proximity)^2} , (1+\Proximity)^2 ]
	\]

    There is a chance that the prover might not declare every relevant bucket of the distribution. 
    The prover gets away with this if none of the verifiers samples are from said relevant bucket. 
    However as $s \Def \frac{10\NumBuckets}{\Proximity}\log(10\NumBuckets)$, using a simple tail bound argument that $L\left(1 - \frac{\Proximity}{10\NumBuckets}\right)^s \le 1/20$, it is easy to see that with probability at least 1 - 1/10, every relevant bucket will be represented in the samples.
	For any $x \in S$, once again we use
    $\yStar(x)$ to denote the corresponding $y_k$ such that $\Claimed{\BucketOf{\Proximity}{x}} = k$.
	We have $|\delta_x| \le \gamma$ for all $x \in S$, where $\delta_x \Def \alpha_{x \yStar} - \frac{\TrueDist{x}}{\TrueDist{\yStar}} $, (Safe to assume \texttt{COMPARE} did not output High or Low, or we would reject immediately), with probability at most 1/20. 
    With a bit of algebra we get

	\begin{equation}
		\ClaimedDist{\yStar(x)}\alpha_{x \yStar(x)} = r(x) \TrueDist{x} + \ClaimedDist{\yStar(x)}\delta_x	\label{eq:algebra}
	\end{equation}

	Thus,
	\begin{align}
		|\ClaimedDist{\yStar(x)}\alpha_{x \yStar(x)} - \ClaimedDist{x}|
		 & = |(\TrueDist{x}-\ClaimedDist{x}) + (r(x)-1)\TrueDist{x} + \ClaimedDist{\yStar(x)}\delta_x| \label{eq:use-algebra}        \\
		 & \ge |\TrueDist{x}-\ClaimedDist{x}| - |r(x) - 1|\,\TrueDist{x} - \ClaimedDist{\yStar(x)}|\delta_x|\label{eq:rev-triangle}
	\end{align}

	Where \eqref{eq:use-algebra} follows from Equation \eqref{eq:algebra} and adding $\TrueDist{x} - \TrueDist{x} = 0$ to the equation, and Inequality \eqref{eq:rev-triangle} comes from the reverse-triangle inequality.
	Summing over $x\in S$ gives, with probability at least $1 - 1/20$,
	\begin{align}
		\sum_{x \in S}\Big|\ClaimedDist{\yStar(x)}\alpha_{x \yStar(x)} - \ClaimedDist{x}\Big|
		 & \ge \sum_{x\in S} \Big|\TrueDist{x}-\ClaimedDist{x}\Big| - \Big|r(x)-1\Big|\sum_{x \in S} \TrueDist{x} - \ClaimedDist{\yStar(x)} \sum_{x \in S}\Big|\delta_x\Big| \\
		 & \ge \sum_{x \in S} \Big|\TrueDist{x}-\ClaimedDist{x}\Big| - |r(x)-1| - \ClaimedDist{\yStar(x)} s\gamma	\label{eq:a}                                               \\
		 & \ge \sum_{x \in S} \Big|\TrueDist{x}-\ClaimedDist{x}\Big| - (2\Proximity + \Proximity^2) - \ClaimedDist{\yStar(x)} s\gamma	\label{eq:b}                            \\
		 & \geq 2\HistFactor - (2\Proximity + \Proximity^2) - \ClaimedDist{\yStar(x)} s\gamma \label{eq:tv}                                                                   \\
		 & \geq 2\HistFactor - (2\Proximity + \Proximity^2) - \frac{\Proximity}{1+\Proximity}
	\end{align}

	Inequality \eqref{eq:a} comes from $\sum_{x\in S} \TrueDist{x} \le 1$ and $\sum_{x \in S}|\delta_x|\le s\gamma$.
	Inequality \eqref{eq:b} comes from the assumption on $r$
	\[
		|r(x)-1| \le \max\!\left\{(1+\Proximity)^2-1,\;1-\frac{1}{(1+\Proximity)^2}\right\}
		= 2\Proximity+\Proximity^2.
	\]

	Inequality \eqref{eq:tv} is a direct consequence of Lemma~\ref{lemma:rel-rl-emd}. 
  Indeed, observe that $\sum_{x \in S} \Big|\TrueDist{x}-\ClaimedDist{x}\Big| = \EMD{\TrueHist}{\ClaimHistExpanded}$ denotes the minimal probability mass of dirt one must move from $\TrueHist$ to construct $\ClaimHistExpanded$ as  $\Claimed{\Dist} =\text{arg}\min_{D' \leftarrow \ClaimHistExpanded} \TV{\True{\Dist}}{D'}$.\\

  \textbf{Complexity}: The verifier needs to run  $\texttt{ApproximateSingle}$ at most $\NumBuckets$ times. 
        This give us a query and sample complexity of $\Tilde{O}\left(\frac{(\log \DomainSize)^4}{\poly(\Proximity)}\right)$.
      As every thing is done by sequential repetition, the final round complexity is $
    \BigOTilde{\log \DomainSize \cdot \poly(1/\Proximity)}$.
    As the communication of each $\texttt{ApproximateSingle}$ is at most $\BigOTilde{\sqrt{\DomainSize}}$, by repeating $\NumBuckets = \BigO{\log \DomainSize}$ times, we get a $\BigOTilde{\sqrt{\DomainSize}}$ communication complexity.

\end{proof}

\section{Conclusion And Open Problems}

We have shown that \emph{any} label-invariant property of a distribution can be efficiently certified using $\poly(\log\DomainSize)$ pairwise conditional queries, and with sublinear communication.
In this section, we outline some open problems and future directions.
\begin{enumerate}
    \item Is it possible to obtain \emph{constant} (or, as a perhaps more achievable goal, depending only on $\Proximity$ and not on $\DomainSize$) query complexity for verifying any label-invariant property with sub-linear communication? This question applies to the stronger $\Cond$ model as well.
    \item Although we achieve $\poly(\log \DomainSize)$ complexity, there is still a gap between the identity tester that makes $\BigO{\sqrt{\log \DomainSize}}$ queries, and our verifier which makes $\BigO{\log^4 \DomainSize}$ queries. Can we close this gap, or is this is an inherent artefact of having less information, given the requirement of sub-linear communication? 
    \item Our protocol currently requires $\poly(\log \DomainSize)$ rounds of interaction. The main culprit here is sequential composition. Are these protocols composable in parallel? This would immediately give us constant-round communication. \citep{herman2022verifying} give a 2-round interactive proof. 
    Given that our support size protocol requires amplification that depends on the parameter $\Proximity$, we cannot hope to achieve universally constant round complexity using just the techniques described in this paper.
    \item Currently, our analysis requires the prover to have full (exact) knowledge of the underlying probability distribution. \citet{herman2023doubley} introduce ``doubly-efficient'' proofs, where the prover itself only needs to learn the distribution up to small error. Can we tighten the analysis, or modify the protocol to accommodate this weaker version of the prover?
    We conjecture this is possible by tightening the analysis of our protocols.
    
    \item Very little is known about the minimum amount of information a prover must communicate for efficient verification in both the sample and conditional access models. As a concrete question, can the $O(\sqrt{\DomainSize})$ communication be improved for the broad class of label-invariant properties?
\end{enumerate}

\bibliographystyle{abbrvnat}
\bibliography{ipdt}
\appendix

\end{document}